\documentclass[aip, reprint]{revtex4-1}

\draft % marks overfull lines with a black rule on the right

\usepackage{graphicx}% Include figure files
\usepackage[colorlinks = true, linkcolor = cyan]{hyperref}
\usepackage{amsmath, amssymb}
\usepackage{comment}
\usepackage{subfig}
\usepackage{bm}

\DeclareMathOperator*{\argmin}{arg\,min}

\newcommand{\pkg}[1]{\texttt{#1}}

\begin{document}

% Use the \preprint command to place your local institutional report number 
% on the title page in preprint mode.
% Multiple \preprint commands are allowed.
%\preprint{}

\title{Forecasting the Forced van der Pol Equation with Frequent Phase Shifts Using Reservoir Computing} %Title of paper

% repeat the \author .. \affiliation  etc. as needed
% \email, \thanks, \homepage, \altaffiliation all apply to the current author.
% Explanatory text should go in the []'s, 
% actual e-mail address or url should go in the {}'s for \email and \homepage.
% Please use the appropriate macro for the type of information

% \affiliation command applies to all authors since the last \affiliation command. 
% The \affiliation command should follow the other information.

\author{Sho Kuno}
\email{kunosho1225@g.ecc.u-tokyo.ac.jp}
%%\homepage[]{Your web page}
%%\thanks{}
%%\altaffiliation{}
\affiliation{Department of Mathematical Informatics, The University of Tokyo, Tokyo 113-8656, Japan.}

\author{Hiroshi Kori}
 \email{kori@k.u-tokyo.ac.jp}
\affiliation{Department of Mathematical Informatics, The University of Tokyo, Tokyo 113-8656, Japan.}
\affiliation{Department of Complexity Sciences and Engineering, The University of Tokyo, Kashiwa, Chiba 277-8561, Japan.}

% Collaboration name, if desired (requires use of superscriptaddress option in \documentclass). 
% \noaffiliation is required (may also be used with the \author command).
%\collaboration{}
%\noaffiliation

\date{\today}

\begin{abstract}
%我々はReservoir computingによる、特定の非自励システムのダイナミクスの予測性能を検証した。具体的にはvan del pol振動子に対して、頻繁な位相シフトを伴う周期外力を与えた系を考える。ある特定の位相シフトに対して生成されたシミュレーションデータで学習と最適化を行ったRCが、異なる位相シフトを伴う周期外力下での振動ダイナミクスを予測させた。その結果、学習データがある程度複雑であれば、異なる位相シフトにさらされた振動ダイナミクスを定量的に予測できることが示された。この設定は、シフトワーカーの概日リズムの状態を予測し、各個人にベターなシフトワークスケジュールをデザインするという問題に動機づけられており、RCがそのような用途に活かせる可能性を示唆するものである。
We tested the performance of reservoir computing (RC) in predicting the dynamics of a certain non-autonomous dynamical system. Specifically, we considered a van del Pol oscillator subjected to periodic external force with frequent phase shifts. The reservoir computer, which was trained and optimized with simulation data generated for a particular phase shift, was designed to predict the oscillation dynamics under periodic external forces with different phase shifts. The results suggest that if the training data have some complexity, it is possible to quantitatively predict the oscillation dynamics exposed to different phase shifts. The setting of this study was motivated by the problem of predicting the state of the circadian rhythm of shift workers and designing a better shift work schedule for each individual. Our results suggest that RC could be exploited for such applications.
%    Non-autonomous dynamical systems have a wide variety of applications, particularly in the study of circadian rhythms. The effects of shift work and jet lag can be examined by modeling a dynamical system with an external drive representing the light-day cycle. However, creating a reliable model for predicting the behavior of such systems is challenging due to their complex and non-linear nature. To address this, we utilized the Reservoir Computer, a framework of Recurrent Neural Networks known for its high performance in predicting non-linear dynamics, to forecast a non-autonomous system with frequent phase shifts in its external drive. Our results suggest that the RC performs well in the prediction task for non-autonomous dynamical systems with a novel setting. These findings imply that RC can contribute to foreseeing the health condition of shift workers prior to schedule changes.

\end{abstract}

\pacs{}% insert suggested PACS numbers in braces on next line

\maketitle %\maketitle must follow title, authors, abstract and \pacs

\begin{quotation}
%睡眠覚醒のスケジュールを定常的にシフトさせるシフトワーカーは、様々な病気のリスクが増加することが知られているが、その背後に概日リズムが乱れがあると考えられる。明暗サイクルにさらされた概日リズムの挙動は、周期外力をうけるリミットサイクルという非自励システムとしてモデル化されることが一般的である。シフトワーカーが、あるシフトワークスケジュールから別のスケジュールへの変更を選択できることを想定しよう。もし事前に新しいスケジュール下での概日リズムの挙動が予測できれば、その情報を判断にいかせる。数理モデルはこれを可能とする道具である。しかし、信頼性の高い数理モデルを構築することは難しい。さらに、限られた観測データから各個人に最適化されたパラメータを得ることも困難であろう。Reservoir Computer (RC)はこの問題に別の解決方法を提供する可能性がある。本研究は、このような背景から、シフトワーカーが経験するような頻繁な位相シフトを伴う周期外力をうけるリミットサイクル振動子のダイナミクスを、RCが予測できるかを検証した。その結果、RCがシフトワーカーのウェルビーイングを高めることに貢献できる可能性を示唆する結果をえた。
Shift workers who regularly shift their sleep-wake schedules are known to have increased risk of various diseases, and circadian rhythm disturbances are thought to be an important risk factor. The behavior of circadian rhythms exposed to light-dark cycles is commonly modeled as a non-autonomous system, namely, a limit-cycle oscillator subjected to external periodic forces. Suppose that a shift worker can choose to change from one shift work schedule to another. If the dynamics of the circadian rhythm under the new schedule can be predicted in advance, this information can be used to make a scheduling decision. Mathematical modeling is a tool that makes this determination possible. However, it is difficult to construct reliable mathematical models precisely reproducing actual dynamics. Moreover, it is difficult to obtain optimized parameters for each individual from limited observational data. Reservoir computing (RC) may provide another solution to this problem. Against this background, this study was conducted to examine whether RC can predict the dynamics of limit cycle oscillators subjected to periodic forces with frequent phase shifts, such as those experienced by shift workers. Our results suggest that RC may potentially contribute to forecasting the circadian rhythms of shift workers.

    % Non-autonomous dynamical systems have a wide variety of applications, particularly in the study of circadian rhythms. The effects of shift work and jet lag can be examined by modeling a dynamical system with an external drive representing the light-day cycle. However, creating a reliable model for predicting the behavior of such systems is challenging due to their complex and non-linear nature. To address this, we utilized the Reservoir Computer, a framework of Recurrent Neural Networks known for its high performance in predicting non-linear dynamics, to forecast a non-autonomous system with frequent phase shifts in its external drive. Our results suggest that the RC performs well in the prediction task for non-autonomous dynamical systems with a novel setting. These findings imply that RC can contribute to foreseeing the health condition of shift workers prior to schedule changes.
\end{quotation}

% Body of paper goes here. Use proper sectioning commands. 
% References should be done using the \cite, \ref, and \label commands
\section{Introduction}
\label{sec: Introduction}
%Nonautonomous dynamical system 
Nonautonomous dynamical systems are dynamical systems whose evolution is determined by time-variant external 
%drivers
drives and parameter effects. These systems are responsive to external effects and time-varying conditions. Thus, they are utilized in various fields, such as engineering, physics, biomedical science, ecology, climate, and neuroscience, for modeling various phenomena\cite{strogatz2018nonlinear, guckenheimer2013nonlinear}. 

Circadian clock dynamics, a factor in various important academic and societal issues, is typically modeled as a nonautonomous system.
%An important application of RCs for predicting nonautonomous systems is the prediction of circadian rhythms of shift workers.
%The study of the circadian rhythm is a crucial application in the field of biomedical science.
The
%endogenous
circadian clock is a self-sustained oscillator, which produces daily variations in genetic and physiological activities. Moreover, the clock is approximately synchronized with the light-dark (LD) cycle of the environmental day-night rhythms, with a period of 24 h.
%Transition in biological values linked with
%By treating the LD cycle as external drive into the system, the circadyan rhythm is often modeled qualitatively by a nonautonomous dynamical system
%with an external drive
Recently, the effect of jet lag and shift work on the circadian clock has attracted considerable research attention, because these effects are thought to have a significant impact on human health\cite{kettnerCircadianHomeostasisLiver2016, yamaguchiVasopressinSignalInhibition2018, inokawaChronicCircadianMisalignment2020}.
The circadian clock is often modeled as a limit-cycle oscillator subjected to external periodic driving\cite{gonzeEntrainmentChaosModel2000, kurosawaAmplitudeCircadianOscillations2006, koriAcceleratingRecoveryJet2017a, yamaguchiMiceGeneticallyDeficient2013a}. 
By exploiting a mathematical model, we may anticipate the effect of a phase shift to the LD cycle on the circadian clock. However, it is challenging to construct a reliable model and infer the model parameters to fit real data.
%It is therefore worth testing the validity of predictions made by RCs.

%Machine Learning Approach
Recent data-driven trends in machine learning may offer alternative approaches to address this issue. Recurrent neural networks (RNNs), a framework of artificial neural networks (ANNs), are particularly suited for problems originating from dynamical systems. They exploit the past information along with the present to update their hidden states, enabling them to capture the temporal dependencies of the sequential data. 
In particular, the long short-term memory (LSTM) RNN subtype succeeds in addressing the long-term dependency problem and achieves better precision \cite{mohajerinMultistepPredictionDynamic2019, siami-naminiPerformanceLSTMBiLSTM2019, tanLSTMBasedAnomalyDetection2020, wangNewConceptUsing2017, huangReconstructingCoupledTime2020}. Despite its success in the learning tasks of dynamical systems, LTSM has the drawback of long computing time. A traditional challenge of designing these RNN schemes is achieving a better tradeoff balance of the prediction precision and the computational cost of the training algorithms. 

Reservoir Computing (RC), or echo-state network,
%(ESN),
is an innovative framework for RNNs that combines high fidelity in replicating dynamics and efficiency in computation.
%Owing to its simplified structure,
In RC, the input and hidden layers are initialized with 
%a randomly selected matrix, 
randomly selected matrices, 
and only the readout layer requires training\cite{bolltExplainingSurprisingSuccess2021}. Further, RC requires only a linear regression for fitting, and no back-propagation that involves nonlinear computation is necessary. Despite this
%significant streamlined
simple structure that distinguishes RC from previous RNN frameworks,
RC is remarkably effective for dynamical-systems learning tasks
even for 
%RC is most known
%RC is even known to be effective in the learning and predicting tasks of
chaotic systems\cite{luAttractorReconstructionMachine2018, pathakModelFreePredictionLarge2018, pathakUsingMachineLearning2017, vlachasBackpropagationAlgorithmsReservoir2020, grigoryevaLearningStrangeAttractors2023},
whose sensitive dependence on initial conditions makes the prediction task more daunting. 
Moreover, data driven approaches on learning tasks of partially or sparse observed systems have carefully been inspected\cite{goswamiDelayEmbeddedEchoState2023, gottwaldCombiningMachineLearning2021, herzogReconstructingComplexCardiac2021, riberaModelSelectionChaotic2022, riberaModelSelectionChaotic2022, shahiPredictionChaoticTime2022, yeoDatadrivenReconstructionNonlinear2019, youngDeepLearningDelay2023}.

RC has been tested for learning tasks of nonautonomous dynamical systems as well.
Prominent research was reported in Ref. \onlinecite{kongDigitalTwinsNonlinear2022a}, where the authors used RC to predict the future state of various chaotic systems with external drives.
Specifically, they employed a sinusoidal function with a slowly growing amplitude as an external 
%driver
drive.
%Our question is, then, can RC be utilized to proficiently examine the effect of jet lag and shift work to one's circadyan rhythm? In other words, can RC predict the state of a nonautonomous dynamical system where its external drive experiences a frequent and rapid phase shifts?
%どういう結果が得られたかを書く。
It was demonstrated that RC can successfully predict the system's behavior if the external drive is known.
%ベターなようやく方法があれば検討してください。
This scenario is similar to the problem of predicting a shift worker's biological clock. In the case of shift workers, if they are exposed to light during waking hours and darkness during sleeping hours, the external drive corresponding to shift work is known. However, unlike the scenario described in  
Ref. \onlinecite{kongDigitalTwinsNonlinear2022a}, the external drive involves frequent and abrupt phase shifts corresponding to changes in working hours. It is necessary to investigate whether RC can accurately predict dynamics even for external drives with such phase shifts.

%具体的には次のような問題設定をする。ある特定のシフトワークスケジュールで働くシフトワーカーが、ある日を堺に、異なるシフトワークスケジュールに移行することを検討しているとする。ここで、我々は事前にそのスケジュールの変化が体内時計に与える影響を予測したいとする。この状況を単純な数理モデルで模倣する。

Therefore, in this study, we asked whether RC can predict the dynamics of limit-cycle oscillators subjected to an external drive with frequent and abrupt phase shifts. Data were generated by a simple model: 
We employed the van der Pol equation as a limit-cycle oscillator and modeled the external drive is a sinusoidal function.
We found that RC can indeed precisely predict oscillation dynamics under certain situations.
%Our question is, then, wheter RC can be utilized to predict shift workers' circadian rhythms.
%Our study is formulated to address these questions.
%Therefore, in this study, we assumed a situation of a shift worker who has worked for a certain period of time with regular shifts, and experience a change in his shift work schedule. 
%For the base model, we employed a simple oscillator model with a limit cycle. 
%The van der Pol equation model met our need and was adopted for the base model, a sinusoidal function for the external force. 
%RC was trained on and implemented for prediction tasks of the future state of this system in the presence of phase shifts to the external drive. 
%We prepared data with different numbers of observed variable due to the assumption that observables are often limited in reality.
%Data driven approaches on learning tasks of partially or sparse observed systems are closely inpected too\cite{goswamiDelayEmbeddedEchoState2023, gottwaldCombiningMachineLearning2021, herzogReconstructingComplexCardiac2021, riberaModelSelectionChaotic2022, riberaModelSelectionChaotic2022, shahiPredictionChaoticTime2022, yeoDatadrivenReconstructionNonlinear2019, youngDeepLearningDelay2023}. 

%The structure of this paper.
This paper is organized as follows: Section \ref{sec: method} explains the methods used for our experiments and the results. 
Section \ref{sec: Conclusion and Discussion} presents the conclusion, followed by a discussion of the results and possible applications, focusing on the circadian rhythm.

\section{Method and Results}
\label{sec: method}

We assumed that a person who has experienced a certain schedule of shift working will at some point change to a new, different schedule. Our aim was to forecast the dynamics of the circadian clock for the new schedule on the basis of past data and the new schedule. Our concrete procedure is described here.

\subsection{Model and Simulation}
\label{subsec: model}

\begin{figure*}[t]
    \centering
    \includegraphics[width=0.9\textwidth]{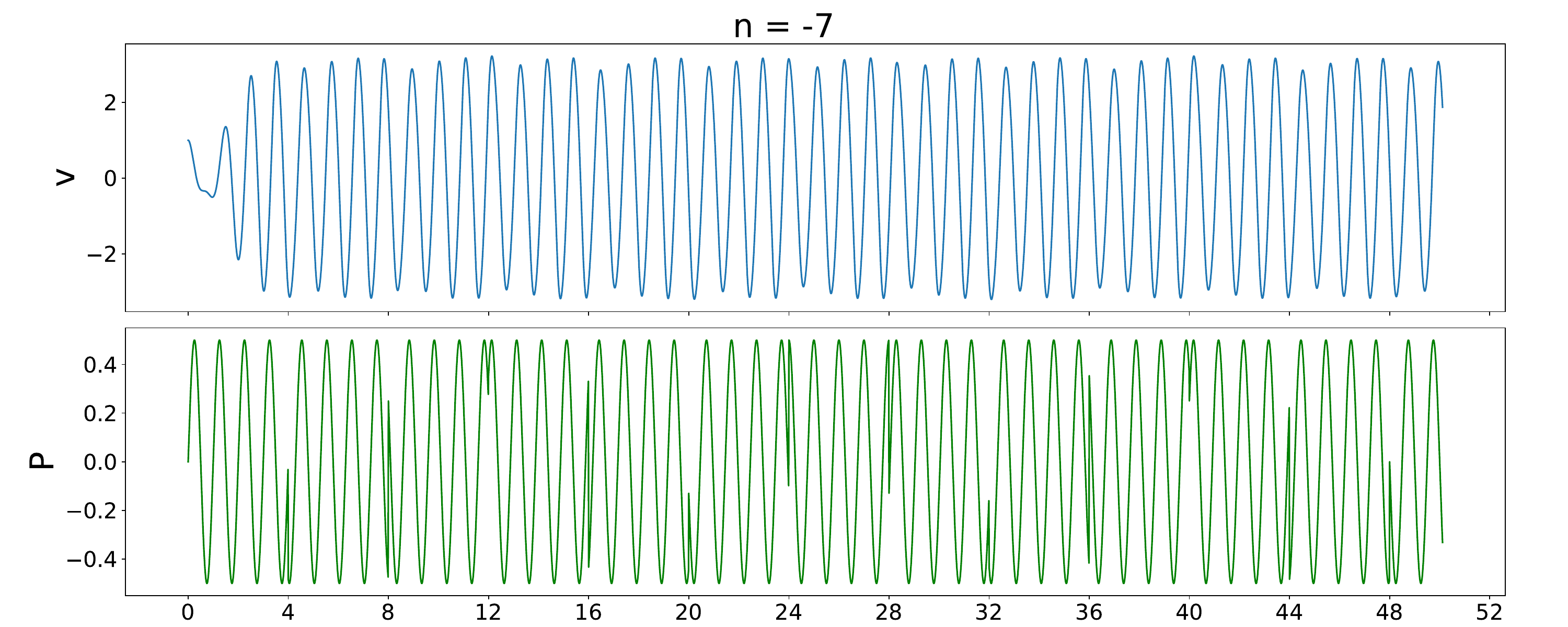}
        \put(-470, 164){(a)}\\
    \includegraphics[width=0.9\textwidth]{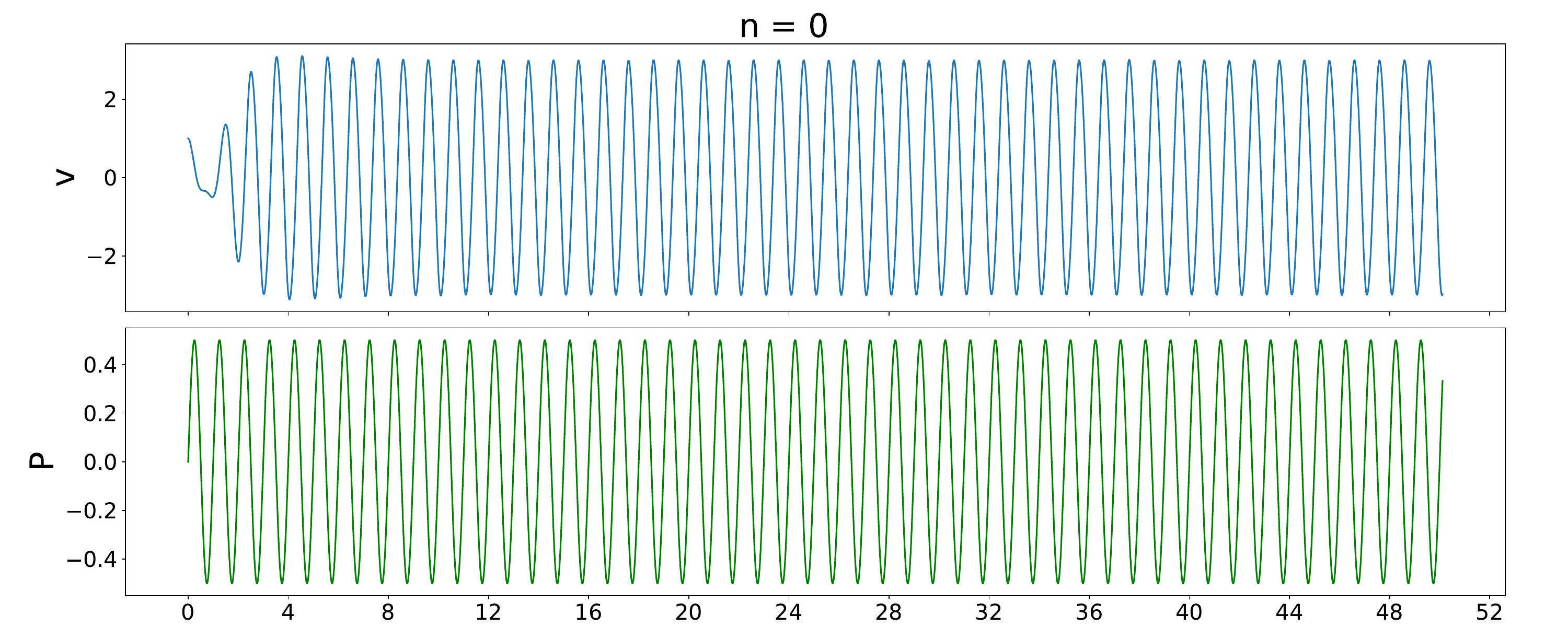}
        \put(-470, 164){(b)}\\
    \includegraphics[width=0.9\textwidth]{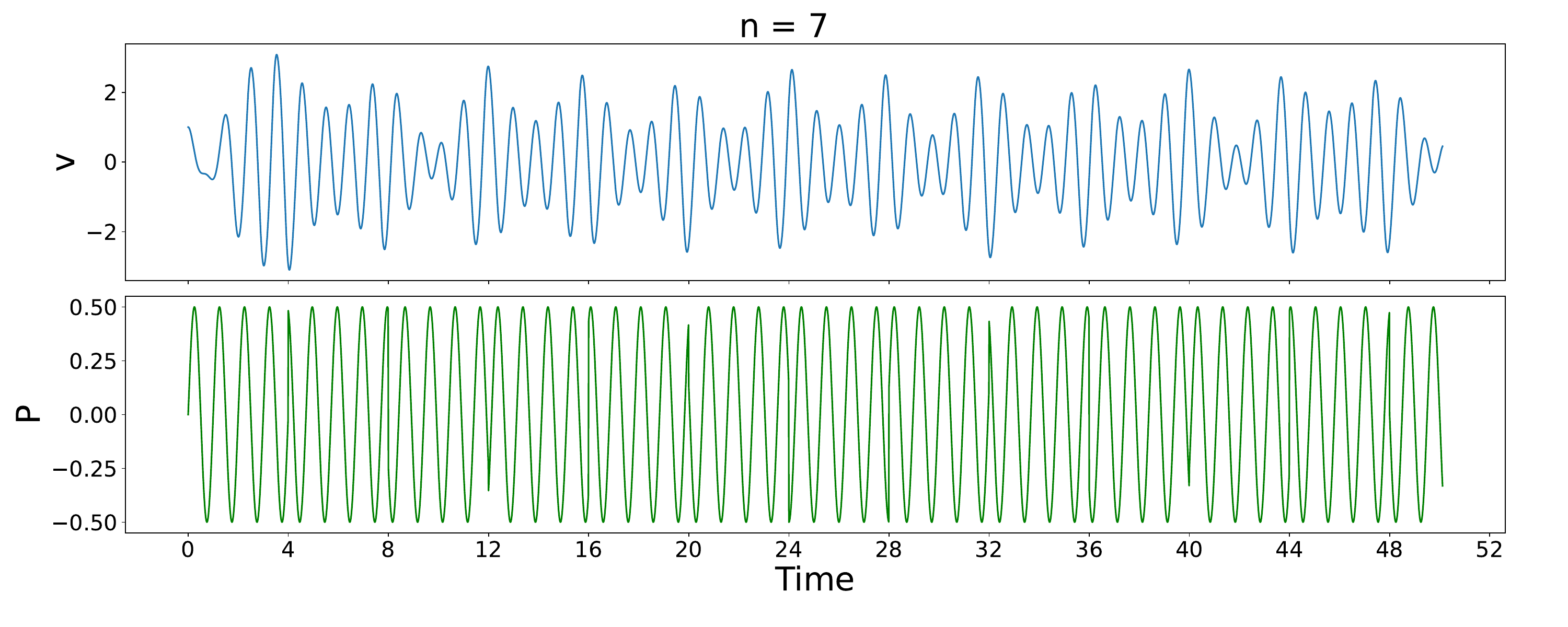}
        \put(-470, 164){(c)}
    \caption{
    Variable $v$ of the van der Pol equation subjected to external drive $P_n(t)$, to which the phase shift of $n$ hours is applied at every 4 d. (a) $n = -7$, (b) $n = 0$, and (c) $n = 7$.\label{Fig: Simulation of Forced VDP}}
\end{figure*}

 We used the forced van der Pol model with an external drive $P_n(t)$, expressed as
\begin{subequations}\label{eq: forced van der Pol}
    \begin{equation}
        \frac{d v}{d t} = w,
    \end{equation}
    \begin{equation}
        \frac{d w}{d t} = \mu(1 - v^2)w - v + P_n(t).
    \end{equation}    
\end{subequations}

%\begin{subequations}\label{eq: forced van der Pol}
    %\begin{equation}
     %   \frac{d x}{d t} = y,
    %\end{equation}
   % \begin{equation}
  %      \frac{d y}{d t} = \mu(1 - x^2)y - x + P_n(t).
 %   \end{equation}    
%\end{subequations}

For $P_n(t)$, we chose the following sinusoidal function with a phase shift function $\theta_n(t)$:
\begin{subequations}
    \begin{equation}
    P_n(t) := A  \sin(\Omega t + \theta_n(t)), \label{eq: external drive}
    \end{equation}
    \begin{equation}
        \theta_n(t) := \frac{n}{24}  \left\lfloor \frac{t}{4 T_e} \right\rfloor  2\pi. 
    \end{equation}
\end{subequations}
where $A$ and $\Omega$ are the strength and frequency of the external drive, respectively, and $T_e := \frac{2\pi}{\Omega}$ is the period of the external drive in the absence of phase shifts. 
The external drive is a model of the day-night rhythm in the context of the circadian clock system; we thus interpreted $T_e$ as one day (1 d). The function $\theta_n(t)$ shifts the phase of $P_n(t)$ by $n$ hours every 4 d, with $n \in \mathbb{Z}, -12 \leq n \leq 12$.

Our numerical simulations were conducted using the \pkg{scipy.integrate.solve$\textunderscore$ivp} package in Python. Time series were sampled with a time step $h := \frac{T_e}{M}$, where $M = 100$ is the number of divisions. To avoid inconveniences caused by the discontinuity of $P_n(t)$ during the simulation, the data were generated by repeating numerical integrations over the individual intervals between each phase shift and concatenating the obtained time series. Because a phase shift was injected every 4 d, each time series segment had $4M$ time steps.
%, ensuring that the entire time series length is a multiple of this duration. 
%The step size of numerical integration is $h := \frac{T_e}{M}$, where $M = 100$ is the number of divisions. To avoid inconveniences caused by the discontinuity of $P_n(t)$ during the simulation, the data are generated by repeating numerical integrations over the individual intervals between each phase shift and concatenating the obtained time series. Because the $P_n(t)$ phase shift is injected every 4 d, each time series segment has $4M$ time steps, ensuring that the entire time series length is a multiple of this duration. 

%For data simulation, various libraries and modules are available in Python. Thus, we used \textbf{scipy.integrate.solve$\textunderscore$ivp}. Other options are available in other languages, including \textbf{ode45} and \textbf{ode15s} in MATLAB.

Throughout the study, we fixed $A = 0.5$ and $\Omega = 1.05$. This choice was motivated by the fact that circadian clocks in mice are considerably disturbed for a phase advancing condition \cite{inokawaChronicCircadianMisalignment2020}. 
Figure \ref{Fig: Simulation of Forced VDP} shows the first 50 d of the data sets (only for the variables $x(t)$ and $P_n(t)$) obtained by a simulation for each $n \in \left\{ -7, 0, 7 \right\}$. The case of $n=0$ corresponds to the forced van der Pol model without any phase shift to the external drive, whereby the obtained time series is genuinely periodic.
For $n=-7$, despite rapid changes in $P_n(t)$, the waveform of $x(t)$ remains nearly periodic. In contrast, for $n=7$, $x(t)$ is considerably distorted and seemingly aperiodic. 
%When comparing the amplitude of $x$,
As targeted, the forced van der Pol equation indeed showed weaker, more disrupted oscillations 
%more weakness
in the forward shifts than the backward ones.

We defined the past schedule of a shift worker as $P_n(t)$ with $n \in \left\{ -7, 0, 7 \right\}$, and the new schedule as $P_{m}(t)$ with $m \in \left\{-11, -10, \cdots, 12 \right\}$. We thus generated the data sets $\{v_i^{(n)}, w_i^{(n)}, P_i^{(n)}\}$
for $n \in \left\{ -7, 0, 7 \right\}$ and divided them into 
two consecutive parts: the training and testing periods, denoted as 
$\{v_{i, \rm train}^{(n)}, w_{i, \rm train}^{(n)}, P_{i, \rm train}^{(n)}\}$ ($i=0,\ldots,N_{\rm train}$) and 
$\{v_{i, \rm test}^{(n)}, w_{i, \rm test}^{(n)}, P_{i, \rm test}^{(n)}\}$ ($i=0,\ldots,N_{\rm test}$), respectively.
As detailed previously, the former and latter were generated for the RC training and the optimization, respectively.
Using the final data as the initial condition, we further generated the data sets for forecast verification, denoted as 
$\{v_{i, \rm forecast}^{(n \to m)}, w_{i, \rm forecast}^{(n \to m)}, P_{i, \rm forecast}^{(m)}\}$ ($i=0,\ldots,N_{\rm forecast}$), where $m \in \left\{-11, -10, \cdots, 12 \right\}$.
We set $N_{\rm train}=8000$, $N_{\rm test}=4000$, $N_{\rm forecast}=3000$, corresponding to the period of 80 d, 40 d, and 30 d.

%We carried out numerical simulations to generate the data sets $\{v_i^{(n)}, w_i^{(n)}, P_i^{(n)}\}$
%for $n=-11,\ldots, +12$ and divided into 
%three consecutive parts: the training, testing, and forecasting periods. 
%Those data are denoted as $v_{i, \rm train}^{(n)}$ ($i=0,\ldots,N_{\rm train}$), $v_{i,\rm test}^{(n)}$ ($i=0,\ldots,N_{\rm test}$), $v_{i, \rm forecast}^{(n)}$ ($i=0,\ldots,N_{\rm forecast}$), and similar notations are applied to $u(t)$ and $P^{(n)}(t)$. We set $N_{\rm train}=8000$, $N_{\rm test}=4000$, $N_{\rm forecast}=3000$, corresponding to the period of 80 d, 40 d, and 30 d.

%Before the experiment, the datasets $\{v_t, w_t, P_t^{(n)}\}$
%%$\mathbf{x}^{n}$
%were divided  The training and testing periods are accessible past information used for training the RC and tuning the hyperparameters. The forecasting period verifies the unknown information about the dynamical system in the future. 

\subsection{Reservoir Computer}\label{subsec: reservoir computer}
\begin{figure*}[t]
    \centering
    \includegraphics[width = \textwidth]{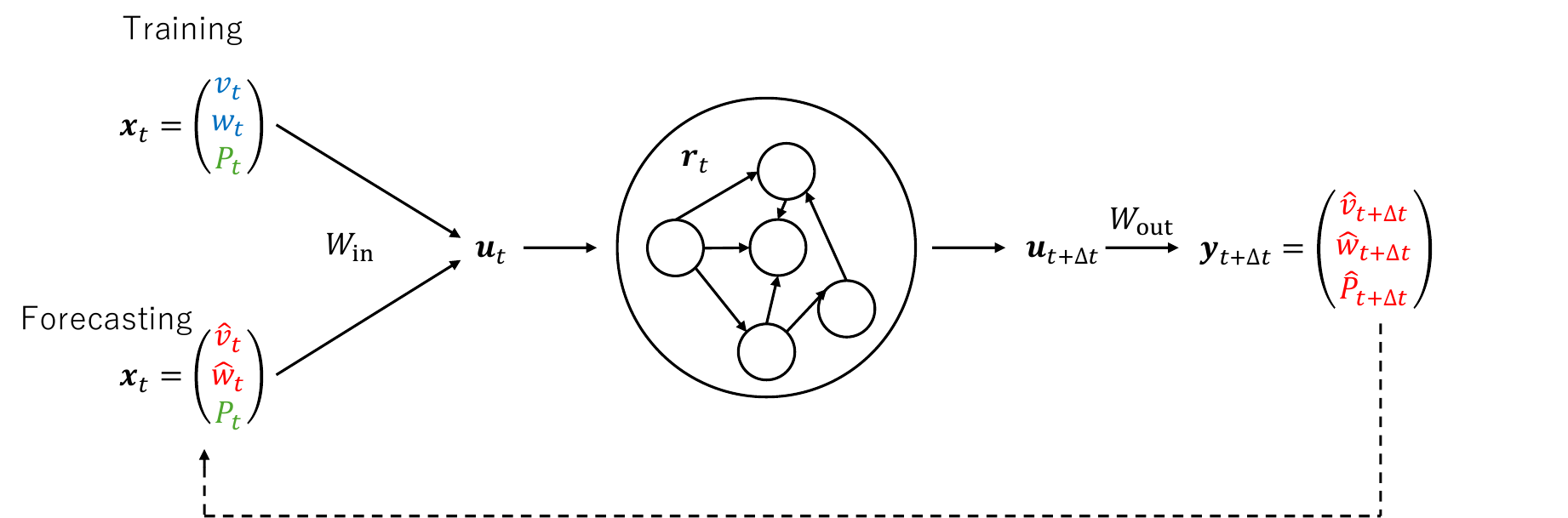}
    \caption{Basic RC structure. The input $\bm{x}_t$ is mapped onto $\bm{u}_t$ in the hidden layer by the matrix $W_{\text{in}}$. During the training and testing periods,
    %within the optimization of hyperparameters,
    the RC is fed with the true data $\bm{x}_t$ of the dynamical system at every step.
    %($v, w$ indicated in blue and $P_n(t)$ in green).
    During the forecasting phase, the reservoir computer updates autonomously using its output as the input for the new step. Here, we inject the true value of the external drive $P_n(t)$ into the input assuming that $P_n(t)$ is accessible at all times, including the forecast period.\label{fig: reservoir computer}}
\end{figure*}

The structure of our RC is illustrated in Fig.~\ref{fig: reservoir computer}.
We adopted the basic RC structure described in Ref. \onlinecite{bolltExplainingSurprisingSuccess2021}. The standard RC model consists of three layers: the input, hidden (reservoir), and readout layers. 
%Let $t$ be the variable for time and
We denoted the data used as the input for updating the reservoir dynamics by
%$\mathbf{z}_t$.
$\bm x_i \in \mathbb R^{d_x}$.
%Define $d_z \in \mathbb{N}$ as the dimension of $\mathbf{z}_t$.
In the input layer, $\bm x_i$ is mapped to a hidden variable $\bm r_i \in \mathbb{R}^{d_r}$ of a much higher dimension $d_r > d_z$ inside the hidden layer by a linear transformation $W_{\rm in}$ of size $d_r \times d_z$:
\begin{eqnarray}
    \bm{u}_i = W_{\rm in} \bm{x}_i.
    \label{eq: input vector}
\end{eqnarray}
In the hidden layer, $\bm r_i$ is updated by a $d_r \times d_r$ linear transformation $W_r$ using an activation function $\tanh(\cdot)$. 
The updating equation of the reservoir state is defined using a variable $\alpha$ of the leaking rate:
\begin{eqnarray}
    \bm{r}_{i+1} = (1-\alpha)\bm{r}_{i} + \alpha \tanh(\bm{u}_{i+1} + W_r \bm{r}_i),
    \label{eq: hidden state evol}
\end{eqnarray}
The leaking rate $\alpha$ is a parameter that controls the effectiveness of the past information on reservoir dynamics, which enables it to regulate the length of the memory the RC can store inside. 
The readout is obtained with another linear transformation $W_{\rm out}$ of size $d_z \times d_r$, 
\begin{eqnarray}
    \bm{y}_i = W_{\rm out}\bm{r}_{i}.
\end{eqnarray}
As a main feature of RC, the matrices ${W}_{\rm in}$ and ${W}_{r}$ are both randomly selected matrices of weights before the training, whereas $W_{\rm out}$ is the only part trained to fit the prediction $\bm{y}_i$ to $\bm{x}_i$.
$W_{\rm out}$ is obtained by solving a least-squares problem of the form 
\begin{eqnarray}
    W_{\rm out} := \underset{V \in \mathbb{R}^{d_x \times d_r}}{\argmin} \|X - V R\|_F,
\end{eqnarray}
where $X=(\bm x_1 \, \bm x_2\, \cdots\, \bm x_{N-1}) \in \mathbb R^{d_x \times {N-1}}$ and $R=(\bm r_1 \, \bm r_2\, \cdots \,\bm r_{N-1}) \in \mathbb R^{d_r \times {N-1}}$.
To solve this problem, we employ Tikhonov regularization with a regularity parameter $\lambda \geq 0$, which yields a formal solution of the form 
\begin{eqnarray}
    {W}_{\rm out} := {Z}{R}^\top\left(RR^\top + \lambda I\right)^{-1},
\end{eqnarray}
where $I$ is the identity matrix.

\subsection{Training and Optimization}\label{subsec: train and optimization}
We used the data sets for $n=-7, 0, +7$ for training and optimizing the reservoir computer.
%Each data set has the length of 12000 (corresponding to 120 d), and they were separated into the training period (the first 80 d) and testing period (the remaining 40 d), denoted as $\mathcal D_n^{\rm train}$ and $\mathcal D_n^{\rm test}$, respectively.
%As the input for the training, we tested two cases: 
%(i) $\bm x_t=(v_t^{(n)({\rm train})}, w_t^{(n)({\rm train})}, P_t^{(n)({\rm train})})$ and (ii) $\bm x_t=(v_t^{(n)({\rm train})}, P_t^{(n)({\rm train})})$. Namely, we use all and partial information for training in the former and latter cases, respectively. 
%In each case, the set $\mathcal D_n^{\rm train}$ was employed to train the RC.
For training, the input $\bm x_i=(v_{i,\rm train}^{(n)}, w_{i,\rm train}^{(n)}, P_{i,\rm train}^{(n)})$ was employed.
After training, the input $\bm x_i=(v_{i,\rm test}^{(n)}, w_{i,\rm test}^{(n)}, P_{i,\rm test}^{(n)})$ was employed for optimizing the reservoir computer. Specifically, the RC hyperparameters were tuned to minimize
%\begin{eqnarray}
%{\rm NRMSE}(\bm{z}, \bm{y}, N)
%{\rm NRMSE}
%= \frac{1}{\lambda}\sqrt{\frac{\sum_{i=0}^{N_{\rm test}-1}\left(\bm{x}_i-\bm{y}_i\right)^2}{N_{\rm test}}},
%\label{eq: objective function}
%\end{eqnarray}

\begin{eqnarray}
%{\rm NRMSE}(\bm{z}, \bm{y}, N)
{\rm NRMSE}
= \sqrt{\frac{\sum_{i=0}^{N_{\rm test}-1}\left(\bm{x}_i-\bm{y}_i\right)^2}{N_{\rm test}}},
\label{eq: objective function}
\end{eqnarray}
%Here, $N$ is the length of the testing period, $\bm{z}$ is the input, and $\bm{y}$ is the prediction by RC during the testing period.
where 
%and
$\bm y_i=(\hat v_i^{(n)}, \hat w_i^{(n)}, \hat P_i^{(n)})$ is the prediction of $\bm x_i$ by the reservoir computer. %$\lambda$ is the factor for normalization, and here we took the difference between the maximum and minimum value of $\bm{x}$ of overall range in the training period.

\begin{figure*}[t]
    \centering
    \includegraphics[width = 0.9\textwidth]{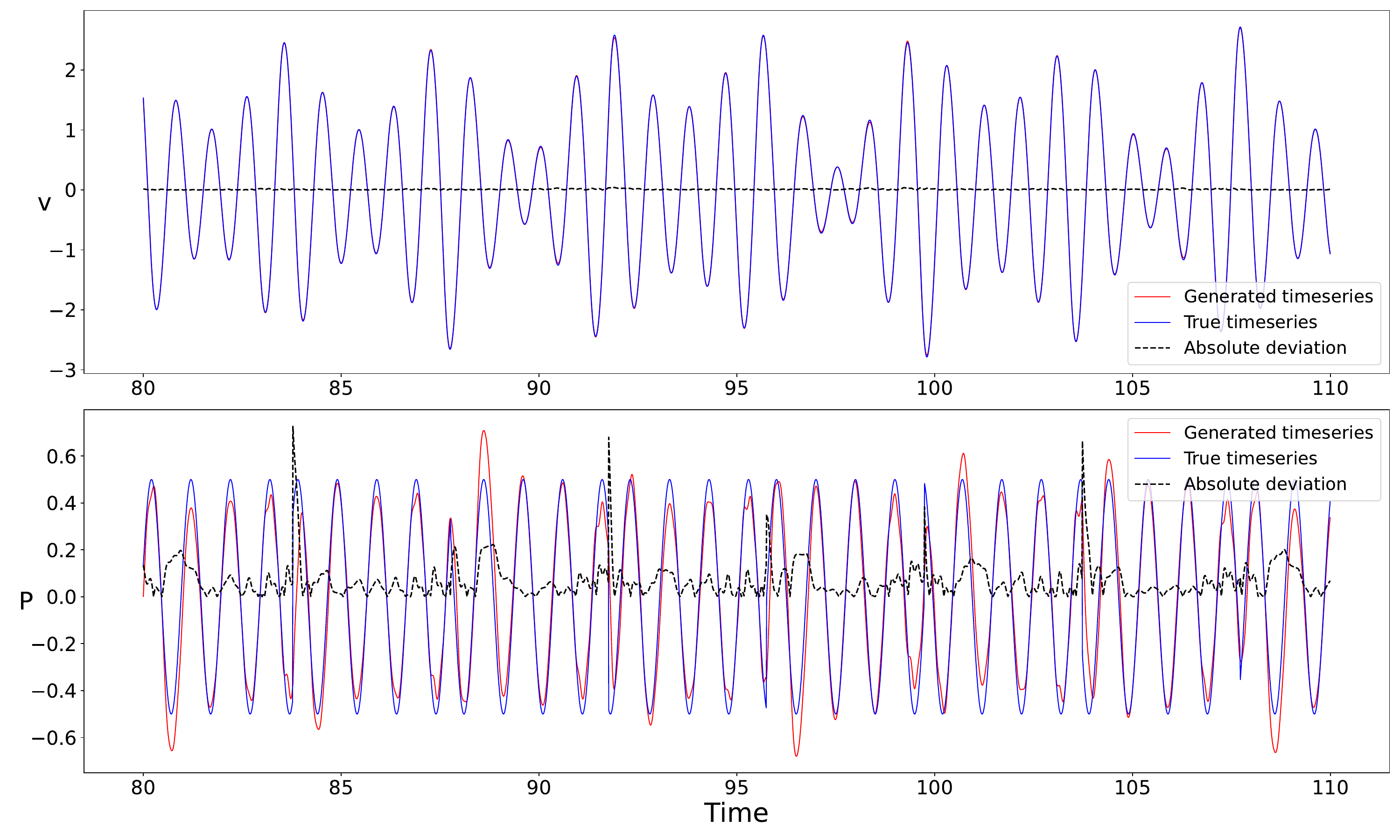}
    \caption{Prediction by RC of the data with $n = 7$ during the testing period. Only $v$ and $P_n(t)$ are shown.
    %from the start of the testing period.
    \label{Fig: prediction for 7}}
\end{figure*}

The obtained hyperparameter sets for each case of $n$ are listed in Appendix \ref{app Hyper-parameters}.
For $n=7$, the comparison between the input and the prediction is shown in Fig. \ref{Fig: prediction for 7}. We observed that the input $v_{i,\rm test}^{(n)}$ and its prediction $\hat v_{i,\rm test}^{(n)}$, which is the component of $\bm y_i$, agree closely. For $n=-7$ and $0$, the input and prediction were also nearly indistinguishable (results not shown).

For the RC implementation, we used the $\pkg{reservoirpy}$ package in Python\cite{trouvainReservoirPyEfficientUserFriendly2020}, with another Python library option available, \pkg{PyRCN}\cite{steinerPyRCNToolboxExploration2022a}. 
For the optimization process, we used \pkg{Optuna}\cite{akibaOptunaNextgenerationHyperparameter2019}, one of the standard libraries for parameter tuning. 
Of the several algorithms available for the optimization and pruning processes, we used \pkg{optuna.samplers.CmaEsSampler} and \pkg{optuna.pruners.SuccessiveHalvingPruner}. 
\pkg{reservoirpy} was designed to be compatible with \pkg{hyperopt}\cite{bergstraHyperoptPythonLibrary2015} as well, but its functions may vary from \pkg{Optuna}. 

\subsection{Forecast}\label{subsec: forecast}
After the RC was trained and optimized for each $n \in \{-7,0,7\}$,
we conducted the forecasting task.
To start forecasting, we “warmed up” the RC with the last insignificant segment in the testing period of $\bm{x}$, contained in the accessible part of the data. Next, we conducted a forecast using 
the input given by
\begin{equation}
 \bm{x}_i=(\hat v_i, \hat w_i, P_{i, \rm forecast}^{(m)}),
\end{equation}
where $m \in \{-11,10,\cdots,12\}$.
Note that, whereas the first two elements are those of the prediction $\bm y_i$, resulting in an input-output loop, the third element is the true external drive.
%We thus tried to forecast the dynamics for the phase shift of $m$ hours under the assumption that only the true drive is accessible. 

\begin{figure*}
\centering
\includegraphics[width = 0.9\textwidth]{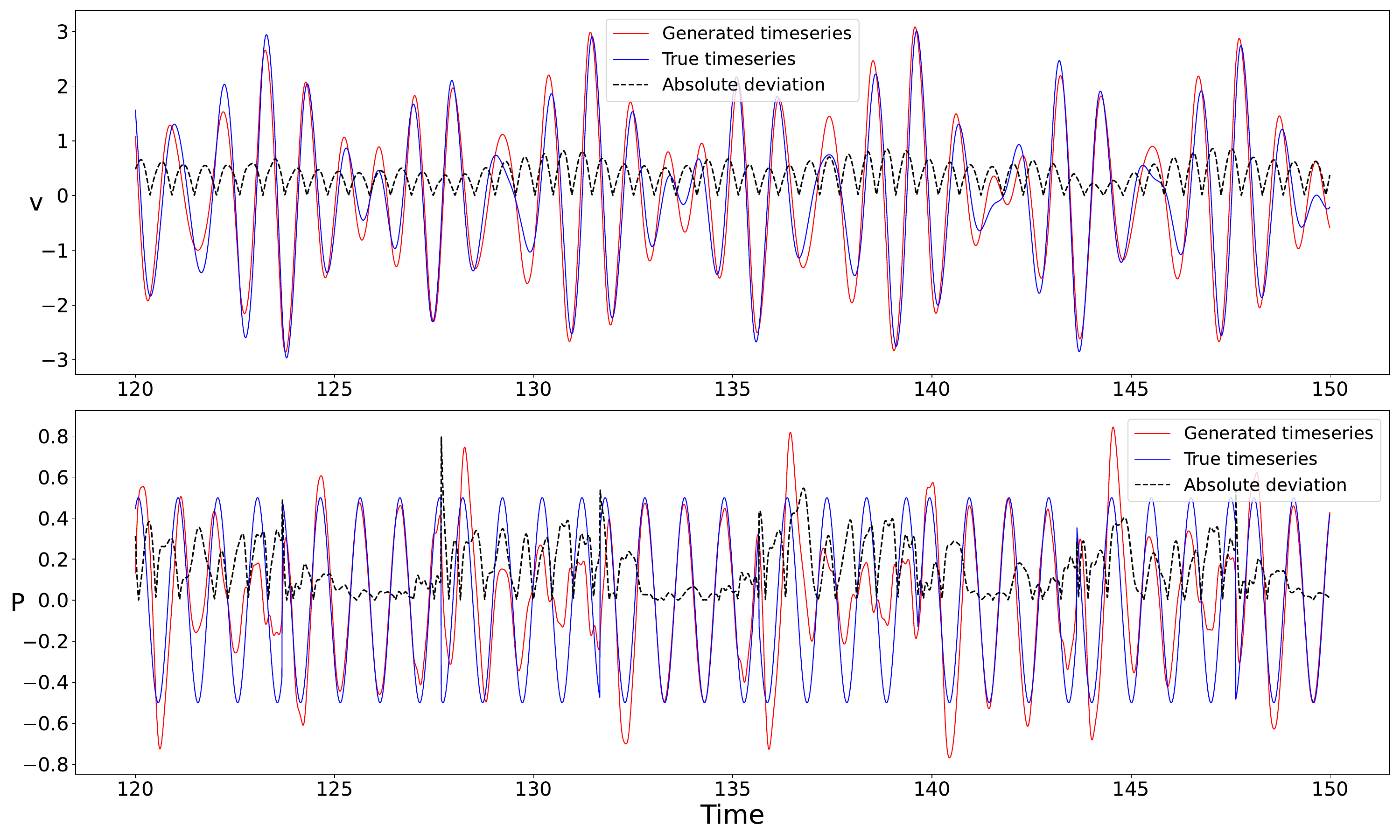}
\caption{Time series of the prediction and true $v_i$ for $n=7$ and $m=10$ during the forecast period. Only $v$ and $P_n(t)$ are shown.}
 \label{fig:forecast}
\end{figure*}

\begin{figure*}
    \centering
    \includegraphics[width = \textwidth]{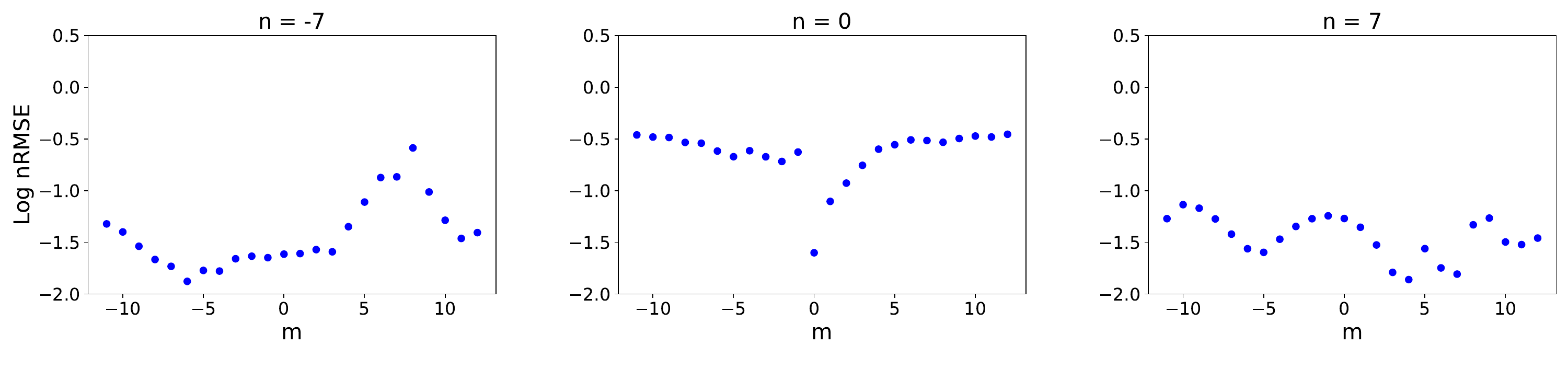}\put(-530, 124){(a)}\\
    \includegraphics[width = \textwidth]{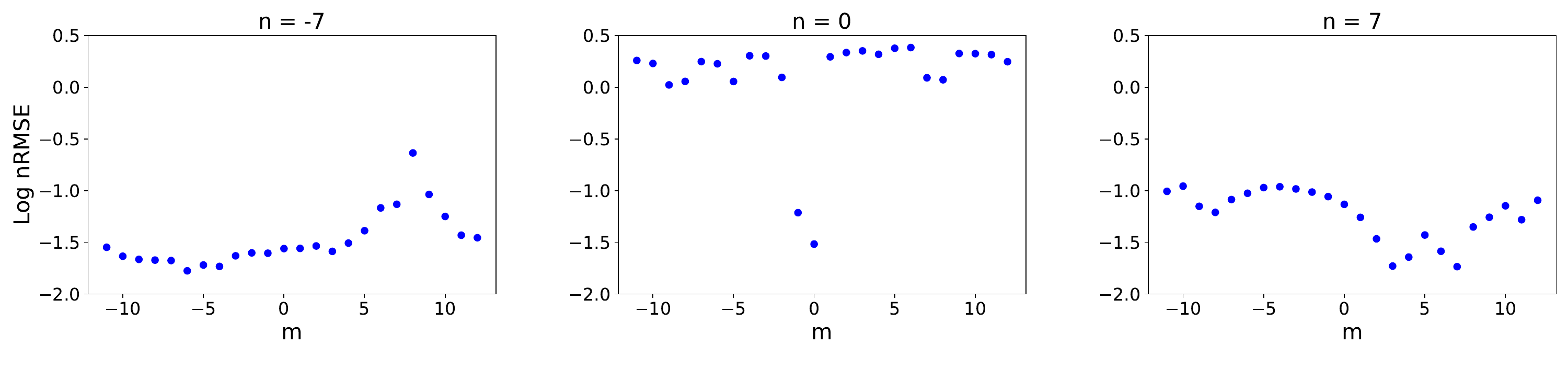}\put(-530, 124){(b)}\\
    \caption{Error of the forecast, $E^{(n\to m)}$, for $n = -7, 0, 7$ and $m \in \{-11,10,\cdots,12\}$,
    where the input was (a) $\bm{x} = (v, w, P_n)$ and 
    (b) $\bm{x} = (v, P_n)$.}
     \label{Fig: 5.forecast_error}
\end{figure*}

An example of the time series of the true and predicted $v_i$ is shown in Fig. \ref{fig:forecast},
where $n=7$ and $m=10$; that is, the training and optimization were done using a shift of $7$ hours and the forecast was conducted for a shift of $10$ hours. 
As shown in Fig.~\ref{Fig: 5.forecast_error}, we quantified the performance of the forecast using the error defined by
\begin{equation}
 E^{(n\to m)} := \log_{10}\sqrt{\frac{\sum_{i=0}^{N_{\rm forecast}-1}\left(\hat v_i - v_{i, \rm forecast}^{(n\to m)}\right)}{N_{\rm forecast}}}.
\end{equation}

%\begin{equation}
 %   E^{(n\to m)} := \frac{1}{\lambda} \sqrt{\frac{\sum_{i=0}^{N_{\rm forecast}-1}\left(\hat v_i - v_{i, \rm forecast}^{(n\to m)}\right)}{N_{\rm forecast}}}, 
%\end{equation}

\begin{figure*}
    \centering
    %\begin{includegraphics}[width=\textwidth]%{stddev/stddev_my_stdev_500_x_y_p_80_40_0.01/std_dev_comparison_my_stdev_500_x_y_p_80_40_0.01.pdf}
    \includegraphics[width = \textwidth]{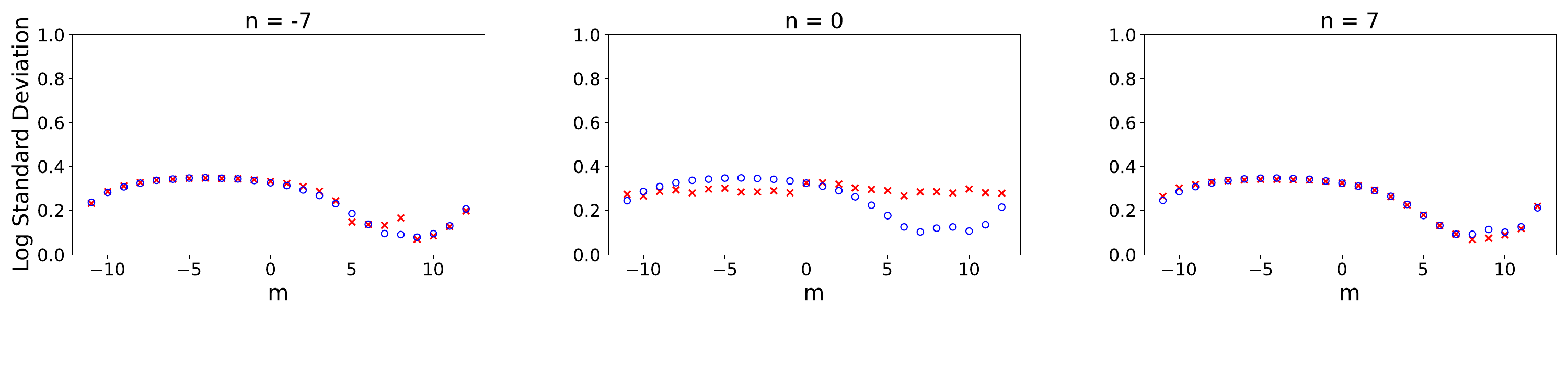}
        \put(-530, 124){(a)}\\
%    \end{includegraphics}
    %\begin{includegraphics}[width=\textwidth]%{stddev/stddev_my_stdev_500_x_p_80_40_0.01/std_dev_comparison_my_stdev_500_x_p_80_40_0.01.pdf}
    \includegraphics[width = \textwidth]{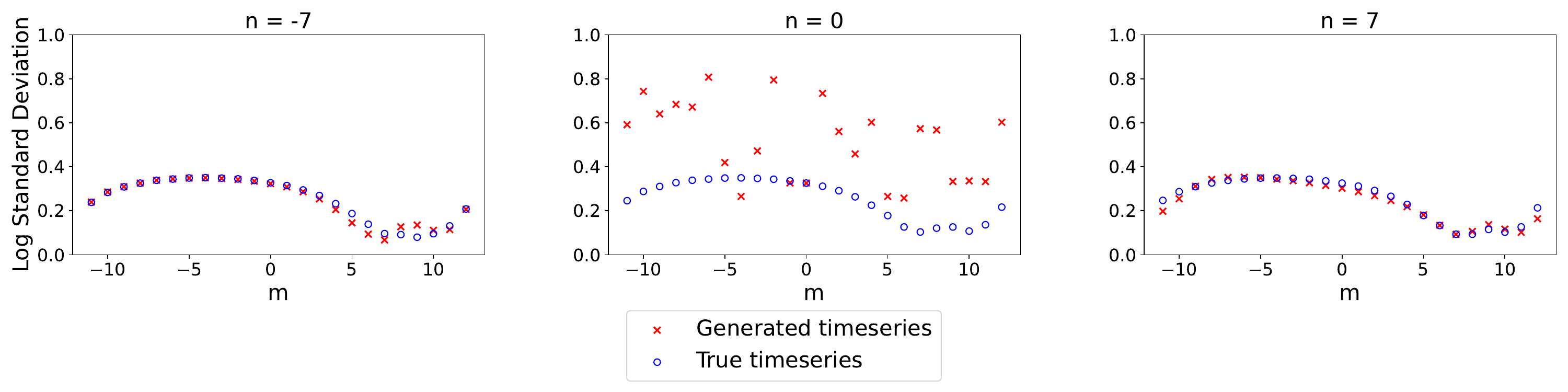}\put(-530, 124){(b)}\\
    %\end{includegraphics}
    \caption{Standard deviation of the true and generated time series of $v(t)$, where the input was
(a)	$\bm{x} = (v, w, P_n)$ and (b) $\bm{x} = (v, P_n)$.}
    \label{Fig: 6. Result xvp8040 v.s. xp8040}
\end{figure*}

For the dataset, we prepared two types, one with array $v, w, P_n(t)$ and the other with $v, P_n(t)$.
For the first type, the reservoir computer was fed with $x, y, P_n(t)$ during the training and testing phases. For the second, only $x, P_n(t)$ were available during the same phase. In Fig. \ref{Fig: 6. Result xvp8040 v.s. xp8040},
we show
%the statistical value of the result and calculated
the standard deviation
%of the amplitude
of the generated and true data for each $n$.
%We took the $\log_{10}$ of the plotted result for each training data shown in Fig. \ref{Fig: 6. Result xvp8040 v.s. xp8040} for the results.

\section{Conclusion and Discussion}\label{sec: Conclusion and Discussion}

In this study, we investigated the forecasting ability of RC using the time series of the forced van der Pol equation with frequent phase shifts to its external drive.
We tested three different sets of training data shown in 
%Fig. 1
Fig. \ref{Fig: Simulation of Forced VDP}; the prediction performance is summarized in Figs. \ref{Fig: 5.forecast_error} and \ref{Fig: 6. Result xvp8040 v.s. xp8040}.
For $n=7$, where the training data were complex (Fig. 1(c)), the predictions were relatively accurate across all values of $m$. For $n=-7$, in which the training data were not highly complex (Fig. 1(a)), the predictions were generally accurate, though the accuracy degraded for some positive values of $m$. For $n=0$, in which the training data were purely periodic (Fig. 1(b)),
the overall results were rather poor, except when $m$ was close to $n=0$. 
Unsurprisingly, the RC performance was excellent when the target data exhibited a pattern similar to that of the training data. It is remarkable that the reservoir computer, trained using a particular data set (i.e., a particular $n$ value), could predict complex dynamics of different situations (i.e., various $m$ values) unless the training data are too simple.
In addition, as shown in Figs. \ref{Fig: 5.forecast_error}(b) and \ref{Fig: 6. Result xvp8040 v.s. xp8040}(b), the reservoir computer could forecast the future even when only some of the variables were observable.
This result is important in applications
where the number of observable variables is limited. 

Our results are consistent with those of Ref. \onlinecite{kongDigitalTwinsNonlinear2022a}, in which it was demonstrated that RC can accurately forecast chaotic dynamical systems with an external drive that has a slowly growing amplitude. In our study, the setting was distinguished from that of this previous study
in that the external drive experienced frequent and rapid phase shifts.
It is generally difficult to predict the response of a system to abrupt changes in perturbations.
From an application point of view, predictions for such perturbations are important and our results support the usefulness of RC.

Finally, we discuss future studies.
As has been noted throughout this paper, the setting of this study was motivated by the problem of predicting shift workers' circadian rhythms. 
Our results show that predictions are possible in a practical setting for simulated data from a simple mathematical model.
However, the predictive power of the proposed model for data generated by more-complex and possibly stochastic models, and more importantly for real data, needs further exploration.

Although we demonstrated the performance of RC with the most basic structure, various advanced versions and RC algorithms exist; options include the online learning\cite{tamuraPartialFORCEFastRobust2021, sussilloGeneratingCoherentPatterns2009} and ensemble methods\cite{shengPredictionIntervalsNoisy2013, yinReservoirComputingEnsembles2012, ortinReservoirComputingEnsemble2017}. The next-generation reservoir computer is another RC scheme that achieves enhanced computational efficiency under certain conditions\cite{gauthierNextGenerationReservoir2021}. 

Theoretical analysis of the prediction of nonautonomous dynamical systems by RC complements our experimental results as well\cite{berryLearningTheoryDynamical2023a}.
Furthermore, RC has been widely studied for aspects of its dimensions\cite{carrollLowDimensionalManifolds2021} and network structures\cite{carrollNetworkStructureEffects2019}. 
The relationship between RC and Taken's embedding theorem is indicated to explain the universal approximation of RC\cite{duanEmbeddingTheoryReservoir2023, grigoryevaChaosCompactManifolds2021, hartEmbeddingApproximationTheorems2020} as well. 
Studies on the role of hyperparameters\cite{stormConstraintsParameterChoices2022, thiedeGradientBasedHyperparameter2019, viehwegParameterizingEchoState2023, zhaoSeekingOptimalParameters2022} provide practical guides on the selection of hyperparameters in various settings. 
These theoretical and practical insights into RC could further enhance its performance in learning nonautonomous dynamical systems, potentially leading to broader applications.

\begin{acknowledgments}
This study was supported by JSPS KAKENHI (No.
JP21K12056, JP22K18384, JP23K27487) to H.K.
\end{acknowledgments}

\section*{Data Availability Statement}
Data available on request from the authors.

\appendix
\label{sec: appendix}
\section{Hyperparameters}
\label{app Hyper-parameters}

The set of hyperparameters subjected to optimization explained in \ref{subsec: train and optimization} are:
\begin{itemize}
    \item \texttt{sr}: the spectral radius of the hidden layer,
    \item \texttt{lr}: leaking rate $\alpha$ that appears in eq. \ref{eq: hidden state evol}. 
    \item \texttt{iss}: the input scaling of $W_{\text{in}}$,
    \item \texttt{ridge}: the ridge value $\lambda$ of ridge regression.
\end{itemize}

The search space of the hyperparameters are:
\[
\begin{aligned}
    \texttt{sr} &\in [10^{-2}, 10], \\
    \texttt{lr} &\in [10^{-9}, 1].\\
    \texttt{iss} &\in [0, 1], \\
    \texttt{ridge} &\in [10^{-9}, 10^{-2}].
\end{aligned}
\]

The set of hyperparameters chosen for the result is presented in Table \ref{table:study_params}. We fixed the cell number of RC to 500.

\begin{table}[h]
\centering
\begin{tabular}{|l|c||c|c|c|c|}
\hline
Variables of Data& n & sr & lr & Iss & ridge\\
\hline
$v, P_n(t)$ & -7 &0.0182 & 0.0583 & 0.6833 & 5.44e-07 \\
\hline
$v, P_n(t)$ & 0 & 0.6338 & 0.4487 & 0.6650 & 2.40e-04 \\
\hline
$v, P_n(t)$ & 7 & 0.0696 & 0.2888 & 0.3732 & 4.69e-07 \\
\hline
$v, w, P_n(t)$ & -7 & 0.3625 & 0.0734 & 0.3330 & 2.96e-05 \\
\hline
$v, w, P_n(t)$ & 0 & 0.2420 & 0.0409 & 0.6981 & 3.86e-07 \\
\hline
$v, w, P_n(t)$ & 7 & 0.0610 & 0.0242 & 0.5831 & 7.67e-03 \\
\hline
\end{tabular}
\caption{Hyperparamters in \ref{subsec: train and optimization}.}
\label{table:study_params}
\end{table}

\bibliography{kuno_0621_edit}% Produces the bibliography via BibTeX.

%merlin.mbs aipnum4-1.bst 2010-07-25 4.21a (PWD, AO, DPC) hacked
%Control: key (0)
%Control: author (8) initials jnrlst
%Control: editor formatted (1) identically to author
%Control: production of article title (0) allowed
%Control: page (1) range
%Control: year (1) truncated
%Control: production of eprint (0) enabled
\begin{thebibliography}{48}%
\makeatletter
\providecommand \@ifxundefined [1]{%
 \@ifx{#1\undefined}
}%
\providecommand \@ifnum [1]{%
 \ifnum #1\expandafter \@firstoftwo
 \else \expandafter \@secondoftwo
 \fi
}%
\providecommand \@ifx [1]{%
 \ifx #1\expandafter \@firstoftwo
 \else \expandafter \@secondoftwo
 \fi
}%
\providecommand \natexlab [1]{#1}%
\providecommand \enquote  [1]{``#1''}%
\providecommand \bibnamefont  [1]{#1}%
\providecommand \bibfnamefont [1]{#1}%
\providecommand \citenamefont [1]{#1}%
\providecommand \href@noop [0]{\@secondoftwo}%
\providecommand \href [0]{\begingroup \@sanitize@url \@href}%
\providecommand \@href[1]{\@@startlink{#1}\@@href}%
\providecommand \@@href[1]{\endgroup#1\@@endlink}%
\providecommand \@sanitize@url [0]{\catcode `\\12\catcode `\$12\catcode `\&12\catcode `\#12\catcode `\^12\catcode `\_12\catcode `\%12\relax}%
\providecommand \@@startlink[1]{}%
\providecommand \@@endlink[0]{}%
\providecommand \url  [0]{\begingroup\@sanitize@url \@url }%
\providecommand \@url [1]{\endgroup\@href {#1}{\urlprefix }}%
\providecommand \urlprefix  [0]{URL }%
\providecommand \Eprint [0]{\href }%
\providecommand \doibase [0]{http://dx.doi.org/}%
\providecommand \selectlanguage [0]{\@gobble}%
\providecommand \bibinfo  [0]{\@secondoftwo}%
\providecommand \bibfield  [0]{\@secondoftwo}%
\providecommand \translation [1]{[#1]}%
\providecommand \BibitemOpen [0]{}%
\providecommand \bibitemStop [0]{}%
\providecommand \bibitemNoStop [0]{.\EOS\space}%
\providecommand \EOS [0]{\spacefactor3000\relax}%
\providecommand \BibitemShut  [1]{\csname bibitem#1\endcsname}%
\let\auto@bib@innerbib\@empty
%</preamble>
\bibitem [{\citenamefont {Strogatz}(2018)}]{strogatz2018nonlinear}%
  \BibitemOpen
  \bibfield  {author} {\bibinfo {author} {\bibfnamefont {S.}~\bibnamefont {Strogatz}},\ }\href@noop {} {\emph {\bibinfo {title} {Nonlinear Dynamics and Chaos: {{With}} Applications to Physics, Biology, Chemistry, and Engineering}}}\ (\bibinfo  {publisher} {CRC Press},\ \bibinfo {year} {2018})\BibitemShut {NoStop}%
\bibitem [{\citenamefont {Guckenheimer}\ and\ \citenamefont {Holmes}(2013)}]{guckenheimer2013nonlinear}%
  \BibitemOpen
  \bibfield  {author} {\bibinfo {author} {\bibfnamefont {J.}~\bibnamefont {Guckenheimer}}\ and\ \bibinfo {author} {\bibfnamefont {P.}~\bibnamefont {Holmes}},\ }\href@noop {} {\emph {\bibinfo {title} {Nonlinear Oscillations, Dynamical Systems, and Bifurcations of Vector Fields}}},\ Applied Mathematical Sciences\ (\bibinfo  {publisher} {Springer New York},\ \bibinfo {year} {2013})\BibitemShut {NoStop}%
\bibitem [{\citenamefont {Kettner}\ \emph {et~al.}(2016)\citenamefont {Kettner}, \citenamefont {Voicu}, \citenamefont {Finegold}, \citenamefont {Coarfa}, \citenamefont {Sreekumar}, \citenamefont {Putluri}, \citenamefont {Katchy}, \citenamefont {Lee}, \citenamefont {Moore},\ and\ \citenamefont {Fu}}]{kettnerCircadianHomeostasisLiver2016}%
  \BibitemOpen
  \bibfield  {author} {\bibinfo {author} {\bibfnamefont {N.~M.}\ \bibnamefont {Kettner}}, \bibinfo {author} {\bibfnamefont {H.}~\bibnamefont {Voicu}}, \bibinfo {author} {\bibfnamefont {M.~J.}\ \bibnamefont {Finegold}}, \bibinfo {author} {\bibfnamefont {C.}~\bibnamefont {Coarfa}}, \bibinfo {author} {\bibfnamefont {A.}~\bibnamefont {Sreekumar}}, \bibinfo {author} {\bibfnamefont {N.}~\bibnamefont {Putluri}}, \bibinfo {author} {\bibfnamefont {C.~A.}\ \bibnamefont {Katchy}}, \bibinfo {author} {\bibfnamefont {C.}~\bibnamefont {Lee}}, \bibinfo {author} {\bibfnamefont {D.~D.}\ \bibnamefont {Moore}}, \ and\ \bibinfo {author} {\bibfnamefont {L.}~\bibnamefont {Fu}},\ }\bibfield  {title} {\enquote {\bibinfo {title} {Circadian {{Homeostasis}} of {{Liver Metabolism Suppresses Hepatocarcinogenesis}}},}\ }\href {\doibase 10.1016/j.ccell.2016.10.007} {\bibfield  {journal} {\bibinfo  {journal} {Cancer Cell}\ }\textbf {\bibinfo {volume} {30}},\ \bibinfo {pages} {909--924} (\bibinfo {year} {2016})}\BibitemShut {NoStop}%
\bibitem [{\citenamefont {Yamaguchi}\ and\ \citenamefont {Okamura}(2018)}]{yamaguchiVasopressinSignalInhibition2018}%
  \BibitemOpen
  \bibfield  {author} {\bibinfo {author} {\bibfnamefont {Y.}~\bibnamefont {Yamaguchi}}\ and\ \bibinfo {author} {\bibfnamefont {H.}~\bibnamefont {Okamura}},\ }\bibfield  {title} {\enquote {\bibinfo {title} {Vasopressin {{Signal Inhibition}} in {{Aged Mice Decreases Mortality}} under {{Chronic Jet Lag}}},}\ }\href {\doibase 10.1016/j.isci.2018.06.008} {\bibfield  {journal} {\bibinfo  {journal} {iScience}\ }\textbf {\bibinfo {volume} {5}},\ \bibinfo {pages} {118--122} (\bibinfo {year} {2018})}\BibitemShut {NoStop}%
\bibitem [{\citenamefont {Inokawa}\ \emph {et~al.}(2020)\citenamefont {Inokawa}, \citenamefont {Umemura}, \citenamefont {Shimba}, \citenamefont {Kawakami}, \citenamefont {Koike}, \citenamefont {Tsuchiya}, \citenamefont {Ohashi}, \citenamefont {Minami}, \citenamefont {Cui}, \citenamefont {Asahi}, \citenamefont {Ono}, \citenamefont {Sasawaki}, \citenamefont {Konishi}, \citenamefont {Yoo}, \citenamefont {Chen}, \citenamefont {Teramukai}, \citenamefont {Ikuta},\ and\ \citenamefont {Yagita}}]{inokawaChronicCircadianMisalignment2020}%
  \BibitemOpen
  \bibfield  {author} {\bibinfo {author} {\bibfnamefont {H.}~\bibnamefont {Inokawa}}, \bibinfo {author} {\bibfnamefont {Y.}~\bibnamefont {Umemura}}, \bibinfo {author} {\bibfnamefont {A.}~\bibnamefont {Shimba}}, \bibinfo {author} {\bibfnamefont {E.}~\bibnamefont {Kawakami}}, \bibinfo {author} {\bibfnamefont {N.}~\bibnamefont {Koike}}, \bibinfo {author} {\bibfnamefont {Y.}~\bibnamefont {Tsuchiya}}, \bibinfo {author} {\bibfnamefont {M.}~\bibnamefont {Ohashi}}, \bibinfo {author} {\bibfnamefont {Y.}~\bibnamefont {Minami}}, \bibinfo {author} {\bibfnamefont {G.}~\bibnamefont {Cui}}, \bibinfo {author} {\bibfnamefont {T.}~\bibnamefont {Asahi}}, \bibinfo {author} {\bibfnamefont {R.}~\bibnamefont {Ono}}, \bibinfo {author} {\bibfnamefont {Y.}~\bibnamefont {Sasawaki}}, \bibinfo {author} {\bibfnamefont {E.}~\bibnamefont {Konishi}}, \bibinfo {author} {\bibfnamefont {S.-H.}\ \bibnamefont {Yoo}}, \bibinfo {author} {\bibfnamefont {Z.}~\bibnamefont {Chen}}, \bibinfo {author} {\bibfnamefont {S.}~\bibnamefont {Teramukai}}, \bibinfo {author} {\bibfnamefont {K.}~\bibnamefont {Ikuta}}, \ and\ \bibinfo {author} {\bibfnamefont {K.}~\bibnamefont {Yagita}},\ }\bibfield  {title} {\enquote {\bibinfo {title} {Chronic circadian misalignment accelerates immune senescence and abbreviates lifespan in mice},}\ }\href {\doibase 10.1038/s41598-020-59541-y} {\bibfield  {journal} {\bibinfo  {journal} {Sci Rep}\ }\textbf {\bibinfo {volume} {10}},\ \bibinfo {pages} {2569} (\bibinfo {year} {2020})}\BibitemShut {NoStop}%
\bibitem [{\citenamefont {Gonze}\ and\ \citenamefont {Goldbeter}(2000)}]{gonzeEntrainmentChaosModel2000}%
  \BibitemOpen
  \bibfield  {author} {\bibinfo {author} {\bibfnamefont {D.}~\bibnamefont {Gonze}}\ and\ \bibinfo {author} {\bibfnamefont {A.}~\bibnamefont {Goldbeter}},\ }\bibfield  {title} {\enquote {\bibinfo {title} {Entrainment {{Versus Chaos}} in a {{Model}} for a {{Circadian Oscillator Driven}} by {{Light-Dark Cycles}}},}\ }\href {\doibase 10.1023/A:1026410121183} {\bibfield  {journal} {\bibinfo  {journal} {Journal of Statistical Physics}\ }\textbf {\bibinfo {volume} {101}},\ \bibinfo {pages} {649--663} (\bibinfo {year} {2000})}\BibitemShut {NoStop}%
\bibitem [{\citenamefont {Kurosawa}\ and\ \citenamefont {Goldbeter}(2006)}]{kurosawaAmplitudeCircadianOscillations2006}%
  \BibitemOpen
  \bibfield  {author} {\bibinfo {author} {\bibfnamefont {G.}~\bibnamefont {Kurosawa}}\ and\ \bibinfo {author} {\bibfnamefont {A.}~\bibnamefont {Goldbeter}},\ }\bibfield  {title} {\enquote {\bibinfo {title} {Amplitude of circadian oscillations entrained by 24-h light--dark cycles},}\ }\href {\doibase 10.1016/j.jtbi.2006.03.016} {\bibfield  {journal} {\bibinfo  {journal} {Journal of Theoretical Biology}\ }\textbf {\bibinfo {volume} {242}},\ \bibinfo {pages} {478--488} (\bibinfo {year} {2006})}\BibitemShut {NoStop}%
\bibitem [{\citenamefont {Kori}, \citenamefont {Yamaguchi},\ and\ \citenamefont {Okamura}(2017)}]{koriAcceleratingRecoveryJet2017a}%
  \BibitemOpen
  \bibfield  {author} {\bibinfo {author} {\bibfnamefont {H.}~\bibnamefont {Kori}}, \bibinfo {author} {\bibfnamefont {Y.}~\bibnamefont {Yamaguchi}}, \ and\ \bibinfo {author} {\bibfnamefont {H.}~\bibnamefont {Okamura}},\ }\bibfield  {title} {\enquote {\bibinfo {title} {Accelerating recovery from jet lag: Prediction from a multi-oscillator model and its experimental confirmation in model animals},}\ }\href {\doibase 10.1038/srep46702} {\bibfield  {journal} {\bibinfo  {journal} {Sci Rep}\ }\textbf {\bibinfo {volume} {7}},\ \bibinfo {pages} {46702} (\bibinfo {year} {2017})}\BibitemShut {NoStop}%
\bibitem [{\citenamefont {Yamaguchi}\ \emph {et~al.}(2013)\citenamefont {Yamaguchi}, \citenamefont {Suzuki}, \citenamefont {Mizoro}, \citenamefont {Kori}, \citenamefont {Okada}, \citenamefont {Chen}, \citenamefont {Fustin}, \citenamefont {Yamazaki}, \citenamefont {Mizuguchi}, \citenamefont {Zhang}, \citenamefont {Dong}, \citenamefont {Tsujimoto}, \citenamefont {Okuno}, \citenamefont {Doi},\ and\ \citenamefont {Okamura}}]{yamaguchiMiceGeneticallyDeficient2013a}%
  \BibitemOpen
  \bibfield  {author} {\bibinfo {author} {\bibfnamefont {Y.}~\bibnamefont {Yamaguchi}}, \bibinfo {author} {\bibfnamefont {T.}~\bibnamefont {Suzuki}}, \bibinfo {author} {\bibfnamefont {Y.}~\bibnamefont {Mizoro}}, \bibinfo {author} {\bibfnamefont {H.}~\bibnamefont {Kori}}, \bibinfo {author} {\bibfnamefont {K.}~\bibnamefont {Okada}}, \bibinfo {author} {\bibfnamefont {Y.}~\bibnamefont {Chen}}, \bibinfo {author} {\bibfnamefont {J.-M.}\ \bibnamefont {Fustin}}, \bibinfo {author} {\bibfnamefont {F.}~\bibnamefont {Yamazaki}}, \bibinfo {author} {\bibfnamefont {N.}~\bibnamefont {Mizuguchi}}, \bibinfo {author} {\bibfnamefont {J.}~\bibnamefont {Zhang}}, \bibinfo {author} {\bibfnamefont {X.}~\bibnamefont {Dong}}, \bibinfo {author} {\bibfnamefont {G.}~\bibnamefont {Tsujimoto}}, \bibinfo {author} {\bibfnamefont {Y.}~\bibnamefont {Okuno}}, \bibinfo {author} {\bibfnamefont {M.}~\bibnamefont {Doi}}, \ and\ \bibinfo {author} {\bibfnamefont {H.}~\bibnamefont {Okamura}},\ }\bibfield  {title} {\enquote {\bibinfo {title} {Mice genetically deficient in vasopressin {{V1a}} and {{V1b}} receptors are resistant to jet lag},}\ }\href {\doibase 10.1126/science.1238599} {\bibfield  {journal} {\bibinfo  {journal} {Science}\ }\textbf {\bibinfo {volume} {342}},\ \bibinfo {pages} {85--90} (\bibinfo {year} {2013})}\BibitemShut {NoStop}%
\bibitem [{\citenamefont {Mohajerin}\ and\ \citenamefont {Waslander}(2019)}]{mohajerinMultistepPredictionDynamic2019}%
  \BibitemOpen
  \bibfield  {author} {\bibinfo {author} {\bibfnamefont {N.}~\bibnamefont {Mohajerin}}\ and\ \bibinfo {author} {\bibfnamefont {S.~L.}\ \bibnamefont {Waslander}},\ }\bibfield  {title} {\enquote {\bibinfo {title} {Multistep {{Prediction}} of {{Dynamic Systems With Recurrent Neural Networks}}},}\ }\href {\doibase 10.1109/TNNLS.2019.2891257} {\bibfield  {journal} {\bibinfo  {journal} {IEEE Transactions on Neural Networks and Learning Systems}\ }\textbf {\bibinfo {volume} {30}},\ \bibinfo {pages} {3370--3383} (\bibinfo {year} {2019})}\BibitemShut {NoStop}%
\bibitem [{\citenamefont {{Siami-Namini}}, \citenamefont {Tavakoli},\ and\ \citenamefont {Namin}(2019)}]{siami-naminiPerformanceLSTMBiLSTM2019}%
  \BibitemOpen
  \bibfield  {author} {\bibinfo {author} {\bibfnamefont {S.}~\bibnamefont {{Siami-Namini}}}, \bibinfo {author} {\bibfnamefont {N.}~\bibnamefont {Tavakoli}}, \ and\ \bibinfo {author} {\bibfnamefont {A.~S.}\ \bibnamefont {Namin}},\ }\bibfield  {title} {\enquote {\bibinfo {title} {The {{Performance}} of {{LSTM}} and {{BiLSTM}} in {{Forecasting Time Series}}},}\ }in\ \href {\doibase 10.1109/BigData47090.2019.9005997} {\emph {\bibinfo {booktitle} {2019 {{IEEE International Conference}} on {{Big Data}} ({{Big Data}})}}}\ (\bibinfo {year} {2019})\ pp.\ \bibinfo {pages} {3285--3292}\BibitemShut {NoStop}%
\bibitem [{\citenamefont {Tan}\ \emph {et~al.}(2020)\citenamefont {Tan}, \citenamefont {Hu}, \citenamefont {Zhang}, \citenamefont {Zheng}, \citenamefont {Davis},\ and\ \citenamefont {Park}}]{tanLSTMBasedAnomalyDetection2020}%
  \BibitemOpen
  \bibfield  {author} {\bibinfo {author} {\bibfnamefont {Y.}~\bibnamefont {Tan}}, \bibinfo {author} {\bibfnamefont {C.}~\bibnamefont {Hu}}, \bibinfo {author} {\bibfnamefont {K.}~\bibnamefont {Zhang}}, \bibinfo {author} {\bibfnamefont {K.}~\bibnamefont {Zheng}}, \bibinfo {author} {\bibfnamefont {E.~A.}\ \bibnamefont {Davis}}, \ and\ \bibinfo {author} {\bibfnamefont {J.~S.}\ \bibnamefont {Park}},\ }\bibfield  {title} {\enquote {\bibinfo {title} {{{LSTM-Based Anomaly Detection}} for {{Non-Linear Dynamical System}}},}\ }\href {\doibase 10.1109/ACCESS.2020.2999065} {\bibfield  {journal} {\bibinfo  {journal} {IEEE Access}\ }\textbf {\bibinfo {volume} {8}},\ \bibinfo {pages} {103301--103308} (\bibinfo {year} {2020})}\BibitemShut {NoStop}%
\bibitem [{\citenamefont {Wang}(2017)}]{wangNewConceptUsing2017}%
  \BibitemOpen
  \bibfield  {author} {\bibinfo {author} {\bibfnamefont {Y.}~\bibnamefont {Wang}},\ }\bibfield  {title} {\enquote {\bibinfo {title} {A new concept using {{LSTM Neural Networks}} for dynamic system identification},}\ }in\ \href {\doibase 10.23919/ACC.2017.7963782} {\emph {\bibinfo {booktitle} {2017 {{American Control Conference}} ({{ACC}})}}}\ (\bibinfo {year} {2017})\ pp.\ \bibinfo {pages} {5324--5329}\BibitemShut {NoStop}%
\bibitem [{\citenamefont {Huang}, \citenamefont {Yang},\ and\ \citenamefont {Fu}(2020)}]{huangReconstructingCoupledTime2020}%
  \BibitemOpen
  \bibfield  {author} {\bibinfo {author} {\bibfnamefont {Y.}~\bibnamefont {Huang}}, \bibinfo {author} {\bibfnamefont {L.}~\bibnamefont {Yang}}, \ and\ \bibinfo {author} {\bibfnamefont {Z.}~\bibnamefont {Fu}},\ }\bibfield  {title} {\enquote {\bibinfo {title} {Reconstructing coupled time series in climate systems using three kinds of machine-learning methods},}\ }\href {\doibase 10.5194/esd-11-835-2020} {\bibfield  {journal} {\bibinfo  {journal} {Earth System Dynamics}\ }\textbf {\bibinfo {volume} {11}},\ \bibinfo {pages} {835--853} (\bibinfo {year} {2020})}\BibitemShut {NoStop}%
\bibitem [{\citenamefont {Bollt}(2021)}]{bolltExplainingSurprisingSuccess2021}%
  \BibitemOpen
  \bibfield  {author} {\bibinfo {author} {\bibfnamefont {E.}~\bibnamefont {Bollt}},\ }\bibfield  {title} {\enquote {\bibinfo {title} {On explaining the surprising success of reservoir computing forecaster of chaos? {{The}} universal machine learning dynamical system with contrast to {{VAR}} and {{DMD}}},}\ }\href {\doibase 10.1063/5.0024890} {\bibfield  {journal} {\bibinfo  {journal} {Chaos: An Interdisciplinary Journal of Nonlinear Science}\ }\textbf {\bibinfo {volume} {31}},\ \bibinfo {pages} {013108} (\bibinfo {year} {2021})}\BibitemShut {NoStop}%
\bibitem [{\citenamefont {Lu}, \citenamefont {Hunt},\ and\ \citenamefont {Ott}(2018)}]{luAttractorReconstructionMachine2018}%
  \BibitemOpen
  \bibfield  {author} {\bibinfo {author} {\bibfnamefont {Z.}~\bibnamefont {Lu}}, \bibinfo {author} {\bibfnamefont {B.~R.}\ \bibnamefont {Hunt}}, \ and\ \bibinfo {author} {\bibfnamefont {E.}~\bibnamefont {Ott}},\ }\bibfield  {title} {\enquote {\bibinfo {title} {Attractor reconstruction by machine learning},}\ }\href {\doibase 10.1063/1.5039508} {\bibfield  {journal} {\bibinfo  {journal} {Chaos: An Interdisciplinary Journal of Nonlinear Science}\ }\textbf {\bibinfo {volume} {28}},\ \bibinfo {pages} {061104} (\bibinfo {year} {2018})}\BibitemShut {NoStop}%
\bibitem [{\citenamefont {Pathak}\ \emph {et~al.}(2018)\citenamefont {Pathak}, \citenamefont {Hunt}, \citenamefont {Girvan}, \citenamefont {Lu},\ and\ \citenamefont {Ott}}]{pathakModelFreePredictionLarge2018}%
  \BibitemOpen
  \bibfield  {author} {\bibinfo {author} {\bibfnamefont {J.}~\bibnamefont {Pathak}}, \bibinfo {author} {\bibfnamefont {B.}~\bibnamefont {Hunt}}, \bibinfo {author} {\bibfnamefont {M.}~\bibnamefont {Girvan}}, \bibinfo {author} {\bibfnamefont {Z.}~\bibnamefont {Lu}}, \ and\ \bibinfo {author} {\bibfnamefont {E.}~\bibnamefont {Ott}},\ }\bibfield  {title} {\enquote {\bibinfo {title} {Model-{{Free Prediction}} of {{Large Spatiotemporally Chaotic Systems}} from {{Data}}: {{A Reservoir Computing Approach}}},}\ }\href {\doibase 10.1103/PhysRevLett.120.024102} {\bibfield  {journal} {\bibinfo  {journal} {Phys. Rev. Lett.}\ }\textbf {\bibinfo {volume} {120}},\ \bibinfo {pages} {024102} (\bibinfo {year} {2018})}\BibitemShut {NoStop}%
\bibitem [{\citenamefont {Pathak}\ \emph {et~al.}(2017)\citenamefont {Pathak}, \citenamefont {Lu}, \citenamefont {Hunt}, \citenamefont {Girvan},\ and\ \citenamefont {Ott}}]{pathakUsingMachineLearning2017}%
  \BibitemOpen
  \bibfield  {author} {\bibinfo {author} {\bibfnamefont {J.}~\bibnamefont {Pathak}}, \bibinfo {author} {\bibfnamefont {Z.}~\bibnamefont {Lu}}, \bibinfo {author} {\bibfnamefont {B.~R.}\ \bibnamefont {Hunt}}, \bibinfo {author} {\bibfnamefont {M.}~\bibnamefont {Girvan}}, \ and\ \bibinfo {author} {\bibfnamefont {E.}~\bibnamefont {Ott}},\ }\bibfield  {title} {\enquote {\bibinfo {title} {Using machine learning to replicate chaotic attractors and calculate {{Lyapunov}} exponents from data},}\ }\href {\doibase 10.1063/1.5010300} {\bibfield  {journal} {\bibinfo  {journal} {Chaos: An Interdisciplinary Journal of Nonlinear Science}\ }\textbf {\bibinfo {volume} {27}},\ \bibinfo {pages} {121102} (\bibinfo {year} {2017})}\BibitemShut {NoStop}%
\bibitem [{\citenamefont {Vlachas}\ \emph {et~al.}(2020)\citenamefont {Vlachas}, \citenamefont {Pathak}, \citenamefont {Hunt}, \citenamefont {Sapsis}, \citenamefont {Girvan}, \citenamefont {Ott},\ and\ \citenamefont {Koumoutsakos}}]{vlachasBackpropagationAlgorithmsReservoir2020}%
  \BibitemOpen
  \bibfield  {author} {\bibinfo {author} {\bibfnamefont {P.~R.}\ \bibnamefont {Vlachas}}, \bibinfo {author} {\bibfnamefont {J.}~\bibnamefont {Pathak}}, \bibinfo {author} {\bibfnamefont {B.~R.}\ \bibnamefont {Hunt}}, \bibinfo {author} {\bibfnamefont {T.~P.}\ \bibnamefont {Sapsis}}, \bibinfo {author} {\bibfnamefont {M.}~\bibnamefont {Girvan}}, \bibinfo {author} {\bibfnamefont {E.}~\bibnamefont {Ott}}, \ and\ \bibinfo {author} {\bibfnamefont {P.}~\bibnamefont {Koumoutsakos}},\ }\bibfield  {title} {\enquote {\bibinfo {title} {Backpropagation algorithms and {{Reservoir Computing}} in {{Recurrent Neural Networks}} for the forecasting of complex spatiotemporal dynamics},}\ }\href {\doibase 10.1016/j.neunet.2020.02.016} {\bibfield  {journal} {\bibinfo  {journal} {Neural Networks}\ }\textbf {\bibinfo {volume} {126}},\ \bibinfo {pages} {191--217} (\bibinfo {year} {2020})}\BibitemShut {NoStop}%
\bibitem [{\citenamefont {Grigoryeva}, \citenamefont {Hart},\ and\ \citenamefont {Ortega}(2023)}]{grigoryevaLearningStrangeAttractors2023}%
  \BibitemOpen
  \bibfield  {author} {\bibinfo {author} {\bibfnamefont {L.}~\bibnamefont {Grigoryeva}}, \bibinfo {author} {\bibfnamefont {A.}~\bibnamefont {Hart}}, \ and\ \bibinfo {author} {\bibfnamefont {J.-P.}\ \bibnamefont {Ortega}},\ }\bibfield  {title} {\enquote {\bibinfo {title} {Learning strange attractors with reservoir systems},}\ }\href {\doibase 10.1088/1361-6544/ace492} {\bibfield  {journal} {\bibinfo  {journal} {Nonlinearity}\ }\textbf {\bibinfo {volume} {36}},\ \bibinfo {pages} {4674--4708} (\bibinfo {year} {2023})},\ \Eprint {http://arxiv.org/abs/2108.05024} {arXiv:2108.05024 [cs, eess, math]} \BibitemShut {NoStop}%
\bibitem [{\citenamefont {Goswami}(2023)}]{goswamiDelayEmbeddedEchoState2023}%
  \BibitemOpen
  \bibfield  {author} {\bibinfo {author} {\bibfnamefont {D.}~\bibnamefont {Goswami}},\ }\bibfield  {title} {\enquote {\bibinfo {title} {Delay {{Embedded Echo-State Network}}: {{A Predictor}} for {{Partially Observed Systems}}},}\ }\href {\doibase 10.1016/j.ifacol.2023.10.470} {\bibfield  {journal} {\bibinfo  {journal} {IFAC-PapersOnLine}\ }\bibinfo {series} {22nd {{IFAC World Congress}}},\ \textbf {\bibinfo {volume} {56}},\ \bibinfo {pages} {6826--6832} (\bibinfo {year} {2023})}\BibitemShut {NoStop}%
\bibitem [{\citenamefont {Gottwald}\ and\ \citenamefont {Reich}(2021)}]{gottwaldCombiningMachineLearning2021}%
  \BibitemOpen
  \bibfield  {author} {\bibinfo {author} {\bibfnamefont {G.~A.}\ \bibnamefont {Gottwald}}\ and\ \bibinfo {author} {\bibfnamefont {S.}~\bibnamefont {Reich}},\ }\bibfield  {title} {\enquote {\bibinfo {title} {Combining machine learning and data assimilation to forecast dynamical systems from noisy partial observations},}\ }\href {\doibase 10.1063/5.0066080} {\bibfield  {journal} {\bibinfo  {journal} {Chaos: An Interdisciplinary Journal of Nonlinear Science}\ }\textbf {\bibinfo {volume} {31}},\ \bibinfo {pages} {101103} (\bibinfo {year} {2021})}\BibitemShut {NoStop}%
\bibitem [{\citenamefont {Herzog}\ \emph {et~al.}(2021)\citenamefont {Herzog}, \citenamefont {Zimmermann}, \citenamefont {Abele}, \citenamefont {Luther},\ and\ \citenamefont {Parlitz}}]{herzogReconstructingComplexCardiac2021}%
  \BibitemOpen
  \bibfield  {author} {\bibinfo {author} {\bibfnamefont {S.}~\bibnamefont {Herzog}}, \bibinfo {author} {\bibfnamefont {R.~S.}\ \bibnamefont {Zimmermann}}, \bibinfo {author} {\bibfnamefont {J.}~\bibnamefont {Abele}}, \bibinfo {author} {\bibfnamefont {S.}~\bibnamefont {Luther}}, \ and\ \bibinfo {author} {\bibfnamefont {U.}~\bibnamefont {Parlitz}},\ }\bibfield  {title} {\enquote {\bibinfo {title} {Reconstructing {{Complex Cardiac Excitation Waves From Incomplete Data Using Echo State Networks}} and {{Convolutional Autoencoders}}},}\ }\href@noop {} {\bibfield  {journal} {\bibinfo  {journal} {Frontiers in Applied Mathematics and Statistics}\ }\textbf {\bibinfo {volume} {6}} (\bibinfo {year} {2021})}\BibitemShut {NoStop}%
\bibitem [{\citenamefont {Ribera}\ \emph {et~al.}(2022)\citenamefont {Ribera}, \citenamefont {Shirman}, \citenamefont {Nguyen},\ and\ \citenamefont {Mangan}}]{riberaModelSelectionChaotic2022}%
  \BibitemOpen
  \bibfield  {author} {\bibinfo {author} {\bibfnamefont {H.}~\bibnamefont {Ribera}}, \bibinfo {author} {\bibfnamefont {S.}~\bibnamefont {Shirman}}, \bibinfo {author} {\bibfnamefont {A.~V.}\ \bibnamefont {Nguyen}}, \ and\ \bibinfo {author} {\bibfnamefont {N.~M.}\ \bibnamefont {Mangan}},\ }\bibfield  {title} {\enquote {\bibinfo {title} {Model selection of chaotic systems from data with hidden variables using sparse data assimilation},}\ }\href {\doibase 10.1063/5.0066066} {\bibfield  {journal} {\bibinfo  {journal} {Chaos: An Interdisciplinary Journal of Nonlinear Science}\ }\textbf {\bibinfo {volume} {32}},\ \bibinfo {pages} {063101} (\bibinfo {year} {2022})}\BibitemShut {NoStop}%
\bibitem [{\citenamefont {Shahi}, \citenamefont {Fenton},\ and\ \citenamefont {Cherry}(2022)}]{shahiPredictionChaoticTime2022}%
  \BibitemOpen
  \bibfield  {author} {\bibinfo {author} {\bibfnamefont {S.}~\bibnamefont {Shahi}}, \bibinfo {author} {\bibfnamefont {F.~H.}\ \bibnamefont {Fenton}}, \ and\ \bibinfo {author} {\bibfnamefont {E.~M.}\ \bibnamefont {Cherry}},\ }\bibfield  {title} {\enquote {\bibinfo {title} {Prediction of chaotic time series using recurrent neural networks and reservoir computing techniques: {{A}} comparative study},}\ }\href {\doibase 10.1016/j.mlwa.2022.100300} {\bibfield  {journal} {\bibinfo  {journal} {Mach Learn Appl}\ }\textbf {\bibinfo {volume} {8}},\ \bibinfo {pages} {100300} (\bibinfo {year} {2022})}\BibitemShut {NoStop}%
\bibitem [{\citenamefont {Yeo}(2019)}]{yeoDatadrivenReconstructionNonlinear2019}%
  \BibitemOpen
  \bibfield  {author} {\bibinfo {author} {\bibfnamefont {K.}~\bibnamefont {Yeo}},\ }\bibfield  {title} {\enquote {\bibinfo {title} {Data-driven reconstruction of nonlinear dynamics from sparse observation},}\ }\href {\doibase 10.1016/j.jcp.2019.06.039} {\bibfield  {journal} {\bibinfo  {journal} {Journal of Computational Physics}\ }\textbf {\bibinfo {volume} {395}},\ \bibinfo {pages} {671--689} (\bibinfo {year} {2019})}\BibitemShut {NoStop}%
\bibitem [{\citenamefont {Young}\ and\ \citenamefont {Graham}(2023)}]{youngDeepLearningDelay2023}%
  \BibitemOpen
  \bibfield  {author} {\bibinfo {author} {\bibfnamefont {C.~D.}\ \bibnamefont {Young}}\ and\ \bibinfo {author} {\bibfnamefont {M.~D.}\ \bibnamefont {Graham}},\ }\bibfield  {title} {\enquote {\bibinfo {title} {Deep learning delay coordinate dynamics for chaotic attractors from partial observable data},}\ }\href {\doibase 10.1103/PhysRevE.107.034215} {\bibfield  {journal} {\bibinfo  {journal} {Phys. Rev. E}\ }\textbf {\bibinfo {volume} {107}},\ \bibinfo {pages} {034215} (\bibinfo {year} {2023})}\BibitemShut {NoStop}%
\bibitem [{\citenamefont {Kong}\ \emph {et~al.}(2022)\citenamefont {Kong}, \citenamefont {Weng}, \citenamefont {Glaz}, \citenamefont {Haile},\ and\ \citenamefont {Lai}}]{kongDigitalTwinsNonlinear2022a}%
  \BibitemOpen
  \bibfield  {author} {\bibinfo {author} {\bibfnamefont {L.-W.}\ \bibnamefont {Kong}}, \bibinfo {author} {\bibfnamefont {Y.}~\bibnamefont {Weng}}, \bibinfo {author} {\bibfnamefont {B.}~\bibnamefont {Glaz}}, \bibinfo {author} {\bibfnamefont {M.}~\bibnamefont {Haile}}, \ and\ \bibinfo {author} {\bibfnamefont {Y.-C.}\ \bibnamefont {Lai}},\ }\href {\doibase 10.48550/arXiv.2210.06144} {\enquote {\bibinfo {title} {Digital twins of nonlinear dynamical systems},}\ } (\bibinfo {year} {2022}),\ \Eprint {http://arxiv.org/abs/2210.06144} {arXiv:2210.06144 [nlin]} \BibitemShut {NoStop}%
\bibitem [{\citenamefont {Trouvain}\ \emph {et~al.}(2020)\citenamefont {Trouvain}, \citenamefont {Pedrelli}, \citenamefont {Dinh},\ and\ \citenamefont {Hinaut}}]{trouvainReservoirPyEfficientUserFriendly2020}%
  \BibitemOpen
  \bibfield  {author} {\bibinfo {author} {\bibfnamefont {N.}~\bibnamefont {Trouvain}}, \bibinfo {author} {\bibfnamefont {L.}~\bibnamefont {Pedrelli}}, \bibinfo {author} {\bibfnamefont {T.~T.}\ \bibnamefont {Dinh}}, \ and\ \bibinfo {author} {\bibfnamefont {X.}~\bibnamefont {Hinaut}},\ }\bibfield  {title} {\enquote {\bibinfo {title} {{{ReservoirPy}}: {{An Efficient}} and {{User-Friendly Library}} to {{Design Echo State Networks}}},}\ }in\ \href {\doibase 10.1007/978-3-030-61616-8_40} {\emph {\bibinfo {booktitle} {Artificial {{Neural Networks}} and {{Machine Learning}} -- {{ICANN}} 2020}}},\ Vol.\ \bibinfo {volume} {12397},\ \bibinfo {editor} {edited by\ \bibinfo {editor} {\bibfnamefont {I.}~\bibnamefont {Farka{\v s}}}, \bibinfo {editor} {\bibfnamefont {P.}~\bibnamefont {Masulli}}, \ and\ \bibinfo {editor} {\bibfnamefont {S.}~\bibnamefont {Wermter}}}\ (\bibinfo  {publisher} {Springer International Publishing},\ \bibinfo {address} {Cham},\ \bibinfo {year} {2020})\ pp.\ \bibinfo {pages} {494--505}\BibitemShut {NoStop}%
\bibitem [{\citenamefont {Steiner}\ \emph {et~al.}(2022)\citenamefont {Steiner}, \citenamefont {Jalalvand}, \citenamefont {Stone},\ and\ \citenamefont {Birkholz}}]{steinerPyRCNToolboxExploration2022a}%
  \BibitemOpen
  \bibfield  {author} {\bibinfo {author} {\bibfnamefont {P.}~\bibnamefont {Steiner}}, \bibinfo {author} {\bibfnamefont {A.}~\bibnamefont {Jalalvand}}, \bibinfo {author} {\bibfnamefont {S.}~\bibnamefont {Stone}}, \ and\ \bibinfo {author} {\bibfnamefont {P.}~\bibnamefont {Birkholz}},\ }\href {\doibase 10.1016/j.engappai.2022.104964} {\enquote {\bibinfo {title} {{{PyRCN}}: {{A Toolbox}} for {{Exploration}} and {{Application}} of {{Reservoir Computing Networks}}},}\ } (\bibinfo {year} {2022}),\ \Eprint {http://arxiv.org/abs/2103.04807} {arXiv:2103.04807 [cs]} \BibitemShut {NoStop}%
\bibitem [{\citenamefont {Akiba}\ \emph {et~al.}(2019)\citenamefont {Akiba}, \citenamefont {Sano}, \citenamefont {Yanase}, \citenamefont {Ohta},\ and\ \citenamefont {Koyama}}]{akibaOptunaNextgenerationHyperparameter2019}%
  \BibitemOpen
  \bibfield  {author} {\bibinfo {author} {\bibfnamefont {T.}~\bibnamefont {Akiba}}, \bibinfo {author} {\bibfnamefont {S.}~\bibnamefont {Sano}}, \bibinfo {author} {\bibfnamefont {T.}~\bibnamefont {Yanase}}, \bibinfo {author} {\bibfnamefont {T.}~\bibnamefont {Ohta}}, \ and\ \bibinfo {author} {\bibfnamefont {M.}~\bibnamefont {Koyama}},\ }\href {\doibase 10.48550/arXiv.1907.10902} {\enquote {\bibinfo {title} {Optuna: {{A Next-generation Hyperparameter Optimization Framework}}},}\ } (\bibinfo {year} {2019}),\ \Eprint {http://arxiv.org/abs/1907.10902} {arXiv:1907.10902 [cs, stat]} \BibitemShut {NoStop}%
\bibitem [{\citenamefont {Bergstra}\ \emph {et~al.}(2015)\citenamefont {Bergstra}, \citenamefont {Komer}, \citenamefont {Eliasmith}, \citenamefont {Yamins},\ and\ \citenamefont {Cox}}]{bergstraHyperoptPythonLibrary2015}%
  \BibitemOpen
  \bibfield  {author} {\bibinfo {author} {\bibfnamefont {J.}~\bibnamefont {Bergstra}}, \bibinfo {author} {\bibfnamefont {B.}~\bibnamefont {Komer}}, \bibinfo {author} {\bibfnamefont {C.}~\bibnamefont {Eliasmith}}, \bibinfo {author} {\bibfnamefont {D.}~\bibnamefont {Yamins}}, \ and\ \bibinfo {author} {\bibfnamefont {D.~D.}\ \bibnamefont {Cox}},\ }\bibfield  {title} {\enquote {\bibinfo {title} {Hyperopt: A {{Python}} library for model selection and hyperparameter optimization},}\ }\href {\doibase 10.1088/1749-4699/8/1/014008} {\bibfield  {journal} {\bibinfo  {journal} {Comput. Sci. Discov.}\ }\textbf {\bibinfo {volume} {8}},\ \bibinfo {pages} {014008} (\bibinfo {year} {2015})}\BibitemShut {NoStop}%
\bibitem [{\citenamefont {Tamura}\ and\ \citenamefont {Tanaka}(2021)}]{tamuraPartialFORCEFastRobust2021}%
  \BibitemOpen
  \bibfield  {author} {\bibinfo {author} {\bibfnamefont {H.}~\bibnamefont {Tamura}}\ and\ \bibinfo {author} {\bibfnamefont {G.}~\bibnamefont {Tanaka}},\ }\bibfield  {title} {\enquote {\bibinfo {title} {Partial-{{FORCE}}: {{A}} fast and robust online training method for recurrent neural networks},}\ }in\ \href {\doibase 10.1109/IJCNN52387.2021.9533964} {\emph {\bibinfo {booktitle} {2021 {{International Joint Conference}} on {{Neural Networks}} ({{IJCNN}})}}}\ (\bibinfo {year} {2021})\ pp.\ \bibinfo {pages} {1--8}\BibitemShut {NoStop}%
\bibitem [{\citenamefont {Sussillo}\ and\ \citenamefont {Abbott}(2009)}]{sussilloGeneratingCoherentPatterns2009}%
  \BibitemOpen
  \bibfield  {author} {\bibinfo {author} {\bibfnamefont {D.}~\bibnamefont {Sussillo}}\ and\ \bibinfo {author} {\bibfnamefont {L.~F.}\ \bibnamefont {Abbott}},\ }\bibfield  {title} {\enquote {\bibinfo {title} {Generating {{Coherent Patterns}} of {{Activity}} from {{Chaotic Neural Networks}}},}\ }\href {\doibase 10.1016/j.neuron.2009.07.018} {\bibfield  {journal} {\bibinfo  {journal} {Neuron}\ }\textbf {\bibinfo {volume} {63}},\ \bibinfo {pages} {544--557} (\bibinfo {year} {2009})}\BibitemShut {NoStop}%
\bibitem [{\citenamefont {Sheng}\ \emph {et~al.}(2013)\citenamefont {Sheng}, \citenamefont {Zhao}, \citenamefont {Wang},\ and\ \citenamefont {Leung}}]{shengPredictionIntervalsNoisy2013}%
  \BibitemOpen
  \bibfield  {author} {\bibinfo {author} {\bibfnamefont {C.}~\bibnamefont {Sheng}}, \bibinfo {author} {\bibfnamefont {J.}~\bibnamefont {Zhao}}, \bibinfo {author} {\bibfnamefont {W.}~\bibnamefont {Wang}}, \ and\ \bibinfo {author} {\bibfnamefont {H.}~\bibnamefont {Leung}},\ }\bibfield  {title} {\enquote {\bibinfo {title} {Prediction {{Intervals}} for a {{Noisy Nonlinear Time Series Based}} on a {{Bootstrapping Reservoir Computing Network Ensemble}}},}\ }\href {\doibase 10.1109/TNNLS.2013.2250299} {\bibfield  {journal} {\bibinfo  {journal} {IEEE Transactions on Neural Networks and Learning Systems}\ }\textbf {\bibinfo {volume} {24}},\ \bibinfo {pages} {1036--1048} (\bibinfo {year} {2013})}\BibitemShut {NoStop}%
\bibitem [{\citenamefont {Yin}\ and\ \citenamefont {Meng}(2012)}]{yinReservoirComputingEnsembles2012}%
  \BibitemOpen
  \bibfield  {author} {\bibinfo {author} {\bibfnamefont {J.}~\bibnamefont {Yin}}\ and\ \bibinfo {author} {\bibfnamefont {Y.}~\bibnamefont {Meng}},\ }\bibfield  {title} {\enquote {\bibinfo {title} {Reservoir computing ensembles for multi-object behavior recognition},}\ }in\ \href {\doibase 10.1109/IJCNN.2012.6252531} {\emph {\bibinfo {booktitle} {The 2012 {{International Joint Conference}} on {{Neural Networks}} ({{IJCNN}})}}}\ (\bibinfo {year} {2012})\ pp.\ \bibinfo {pages} {1--8}\BibitemShut {NoStop}%
\bibitem [{\citenamefont {Ort{\'i}n}\ and\ \citenamefont {Pesquera}(2017)}]{ortinReservoirComputingEnsemble2017}%
  \BibitemOpen
  \bibfield  {author} {\bibinfo {author} {\bibfnamefont {S.}~\bibnamefont {Ort{\'i}n}}\ and\ \bibinfo {author} {\bibfnamefont {L.}~\bibnamefont {Pesquera}},\ }\bibfield  {title} {\enquote {\bibinfo {title} {Reservoir {{Computing}} with an {{Ensemble}} of {{Time-Delay Reservoirs}}},}\ }\href {\doibase 10.1007/s12559-017-9463-7} {\bibfield  {journal} {\bibinfo  {journal} {Cogn Comput}\ }\textbf {\bibinfo {volume} {9}},\ \bibinfo {pages} {327--336} (\bibinfo {year} {2017})}\BibitemShut {NoStop}%
\bibitem [{\citenamefont {Gauthier}\ \emph {et~al.}(2021)\citenamefont {Gauthier}, \citenamefont {Bollt}, \citenamefont {Griffith},\ and\ \citenamefont {Barbosa}}]{gauthierNextGenerationReservoir2021}%
  \BibitemOpen
  \bibfield  {author} {\bibinfo {author} {\bibfnamefont {D.~J.}\ \bibnamefont {Gauthier}}, \bibinfo {author} {\bibfnamefont {E.}~\bibnamefont {Bollt}}, \bibinfo {author} {\bibfnamefont {A.}~\bibnamefont {Griffith}}, \ and\ \bibinfo {author} {\bibfnamefont {W.~A.~S.}\ \bibnamefont {Barbosa}},\ }\bibfield  {title} {\enquote {\bibinfo {title} {Next generation reservoir computing},}\ }\href {\doibase 10.1038/s41467-021-25801-2} {\bibfield  {journal} {\bibinfo  {journal} {Nat Commun}\ }\textbf {\bibinfo {volume} {12}},\ \bibinfo {pages} {5564} (\bibinfo {year} {2021})}\BibitemShut {NoStop}%
\bibitem [{\citenamefont {Berry}\ and\ \citenamefont {Das}(2023)}]{berryLearningTheoryDynamical2023a}%
  \BibitemOpen
  \bibfield  {author} {\bibinfo {author} {\bibfnamefont {T.}~\bibnamefont {Berry}}\ and\ \bibinfo {author} {\bibfnamefont {S.}~\bibnamefont {Das}},\ }\bibfield  {title} {\enquote {\bibinfo {title} {Learning {{Theory}} for {{Dynamical Systems}}},}\ }\href {\doibase 10.1137/22M1516865} {\bibfield  {journal} {\bibinfo  {journal} {SIAM J. Appl. Dyn. Syst.}\ }\textbf {\bibinfo {volume} {22}},\ \bibinfo {pages} {2082--2122} (\bibinfo {year} {2023})}\BibitemShut {NoStop}%
\bibitem [{\citenamefont {Carroll}(2021)}]{carrollLowDimensionalManifolds2021}%
  \BibitemOpen
  \bibfield  {author} {\bibinfo {author} {\bibfnamefont {T.~L.}\ \bibnamefont {Carroll}},\ }\bibfield  {title} {\enquote {\bibinfo {title} {Low dimensional manifolds in reservoir computers},}\ }\href {\doibase 10.1063/5.0047006} {\bibfield  {journal} {\bibinfo  {journal} {Chaos: An Interdisciplinary Journal of Nonlinear Science}\ }\textbf {\bibinfo {volume} {31}},\ \bibinfo {pages} {043113} (\bibinfo {year} {2021})}\BibitemShut {NoStop}%
\bibitem [{\citenamefont {Carroll}\ and\ \citenamefont {Pecora}(2019)}]{carrollNetworkStructureEffects2019}%
  \BibitemOpen
  \bibfield  {author} {\bibinfo {author} {\bibfnamefont {T.~L.}\ \bibnamefont {Carroll}}\ and\ \bibinfo {author} {\bibfnamefont {L.~M.}\ \bibnamefont {Pecora}},\ }\bibfield  {title} {\enquote {\bibinfo {title} {Network structure effects in reservoir computers},}\ }\href {\doibase 10.1063/1.5097686} {\bibfield  {journal} {\bibinfo  {journal} {Chaos: An Interdisciplinary Journal of Nonlinear Science}\ }\textbf {\bibinfo {volume} {29}},\ \bibinfo {pages} {083130} (\bibinfo {year} {2019})}\BibitemShut {NoStop}%
\bibitem [{\citenamefont {Duan}\ \emph {et~al.}(2023)\citenamefont {Duan}, \citenamefont {Ying}, \citenamefont {Leng}, \citenamefont {Kurths}, \citenamefont {Lin},\ and\ \citenamefont {Ma}}]{duanEmbeddingTheoryReservoir2023}%
  \BibitemOpen
  \bibfield  {author} {\bibinfo {author} {\bibfnamefont {X.-Y.}\ \bibnamefont {Duan}}, \bibinfo {author} {\bibfnamefont {X.}~\bibnamefont {Ying}}, \bibinfo {author} {\bibfnamefont {S.-Y.}\ \bibnamefont {Leng}}, \bibinfo {author} {\bibfnamefont {J.}~\bibnamefont {Kurths}}, \bibinfo {author} {\bibfnamefont {W.}~\bibnamefont {Lin}}, \ and\ \bibinfo {author} {\bibfnamefont {H.-F.}\ \bibnamefont {Ma}},\ }\bibfield  {title} {\enquote {\bibinfo {title} {Embedding theory of reservoir computing and reducing reservoir network using time delays},}\ }\href {\doibase 10.1103/PhysRevResearch.5.L022041} {\bibfield  {journal} {\bibinfo  {journal} {Phys. Rev. Res.}\ }\textbf {\bibinfo {volume} {5}},\ \bibinfo {pages} {L022041} (\bibinfo {year} {2023})}\BibitemShut {NoStop}%
\bibitem [{\citenamefont {Grigoryeva}, \citenamefont {Hart},\ and\ \citenamefont {Ortega}(2021)}]{grigoryevaChaosCompactManifolds2021}%
  \BibitemOpen
  \bibfield  {author} {\bibinfo {author} {\bibfnamefont {L.}~\bibnamefont {Grigoryeva}}, \bibinfo {author} {\bibfnamefont {A.}~\bibnamefont {Hart}}, \ and\ \bibinfo {author} {\bibfnamefont {J.-P.}\ \bibnamefont {Ortega}},\ }\bibfield  {title} {\enquote {\bibinfo {title} {Chaos on compact manifolds: {{Differentiable}} synchronizations beyond the {{Takens}} theorem},}\ }\href {\doibase 10.1103/PhysRevE.103.062204} {\bibfield  {journal} {\bibinfo  {journal} {Phys. Rev. E}\ }\textbf {\bibinfo {volume} {103}},\ \bibinfo {pages} {062204} (\bibinfo {year} {2021})}\BibitemShut {NoStop}%
\bibitem [{\citenamefont {Hart}, \citenamefont {Hook},\ and\ \citenamefont {Dawes}(2020)}]{hartEmbeddingApproximationTheorems2020}%
  \BibitemOpen
  \bibfield  {author} {\bibinfo {author} {\bibfnamefont {A.}~\bibnamefont {Hart}}, \bibinfo {author} {\bibfnamefont {J.}~\bibnamefont {Hook}}, \ and\ \bibinfo {author} {\bibfnamefont {J.}~\bibnamefont {Dawes}},\ }\bibfield  {title} {\enquote {\bibinfo {title} {Embedding and approximation theorems for echo state networks},}\ }\href {\doibase 10.1016/j.neunet.2020.05.013} {\bibfield  {journal} {\bibinfo  {journal} {Neural Networks}\ }\textbf {\bibinfo {volume} {128}},\ \bibinfo {pages} {234--247} (\bibinfo {year} {2020})}\BibitemShut {NoStop}%
\bibitem [{\citenamefont {Storm}, \citenamefont {Gustavsson},\ and\ \citenamefont {Mehlig}(2022)}]{stormConstraintsParameterChoices2022}%
  \BibitemOpen
  \bibfield  {author} {\bibinfo {author} {\bibfnamefont {L.}~\bibnamefont {Storm}}, \bibinfo {author} {\bibfnamefont {K.}~\bibnamefont {Gustavsson}}, \ and\ \bibinfo {author} {\bibfnamefont {B.}~\bibnamefont {Mehlig}},\ }\bibfield  {title} {\enquote {\bibinfo {title} {Constraints on parameter choices for successful time-series prediction with echo-state networks},}\ }\href {\doibase 10.1088/2632-2153/aca1f6} {\bibfield  {journal} {\bibinfo  {journal} {Mach. Learn.: Sci. Technol.}\ }\textbf {\bibinfo {volume} {3}},\ \bibinfo {pages} {045021} (\bibinfo {year} {2022})}\BibitemShut {NoStop}%
\bibitem [{\citenamefont {Thiede}\ and\ \citenamefont {Parlitz}(2019)}]{thiedeGradientBasedHyperparameter2019}%
  \BibitemOpen
  \bibfield  {author} {\bibinfo {author} {\bibfnamefont {L.~A.}\ \bibnamefont {Thiede}}\ and\ \bibinfo {author} {\bibfnamefont {U.}~\bibnamefont {Parlitz}},\ }\bibfield  {title} {\enquote {\bibinfo {title} {Gradient based hyperparameter optimization in {{Echo State Networks}}},}\ }\href {\doibase 10.1016/j.neunet.2019.02.001} {\bibfield  {journal} {\bibinfo  {journal} {Neural Networks}\ }\textbf {\bibinfo {volume} {115}},\ \bibinfo {pages} {23--29} (\bibinfo {year} {2019})}\BibitemShut {NoStop}%
\bibitem [{\citenamefont {Viehweg}, \citenamefont {Worthmann},\ and\ \citenamefont {M{\"a}der}(2023)}]{viehwegParameterizingEchoState2023}%
  \BibitemOpen
  \bibfield  {author} {\bibinfo {author} {\bibfnamefont {J.}~\bibnamefont {Viehweg}}, \bibinfo {author} {\bibfnamefont {K.}~\bibnamefont {Worthmann}}, \ and\ \bibinfo {author} {\bibfnamefont {P.}~\bibnamefont {M{\"a}der}},\ }\bibfield  {title} {\enquote {\bibinfo {title} {Parameterizing echo state networks for multi-step time series prediction},}\ }\href {\doibase 10.1016/j.neucom.2022.11.044} {\bibfield  {journal} {\bibinfo  {journal} {Neurocomputing}\ }\textbf {\bibinfo {volume} {522}},\ \bibinfo {pages} {214--228} (\bibinfo {year} {2023})}\BibitemShut {NoStop}%
\bibitem [{\citenamefont {Zhao}(2022)}]{zhaoSeekingOptimalParameters2022}%
  \BibitemOpen
  \bibfield  {author} {\bibinfo {author} {\bibfnamefont {B.}~\bibnamefont {Zhao}},\ }\bibfield  {title} {\enquote {\bibinfo {title} {Seeking optimal parameters for achieving a lightweight reservoir computing: {{A}} computational endeavor},}\ }\href {\doibase 10.3934/era.2022152} {\bibfield  {journal} {\bibinfo  {journal} {era}\ }\textbf {\bibinfo {volume} {30}},\ \bibinfo {pages} {3004--3018} (\bibinfo {year} {2022})}\BibitemShut {NoStop}%
\end{thebibliography}%

\end{document}